\documentclass[12pt]{article}

\usepackage[dvips]{graphicx}
\usepackage{epsfig}
\usepackage{amsmath,amsfonts,amssymb,amsthm}
\usepackage{mathrsfs,mathtools}
\usepackage{verbatim}
\usepackage{psfrag}
\usepackage{bm}
\usepackage{bbm}
\usepackage[utf8]{inputenc}
\usepackage[square,comma,sort&compress,numbers]{natbib}
\usepackage{xcolor}
\usepackage{slashed}
\usepackage{upgreek}
\usepackage{enumitem}
\usepackage{float}
\usepackage[T1]{fontenc}
\usepackage{soul}
\usepackage{pgfplots}
\pgfplotsset{compat=1.17}
\usetikzlibrary{external}
\usepackage{marginnote}
\usepackage{array}
\tikzset{
  external/only named=true,
  thick/.style={line width=.5pt},
  approximation/.style={line width=1.2pt},
  numerics/.style={black, dotted, line width=.8pt},
  amplitude/.style={dashed},
  estimate/.style={dashed, line width=.8pt},
  normal plot/.style={line width=.8pt},
}

\usepackage[colorlinks,citecolor=black,urlcolor=black,linkcolor=black]{hyperref}

\usepackage{epsf,epsfig}
\usepackage{graphics}
\usepackage{subcaption}

\newcommand\blfootnote[1]{%
  \begingroup
  \renewcommand\thefootnote{}\footnote{#1}%
  \addtocounter{footnote}{-1}%
  \endgroup
}

\def\d{\mathrm{d}}

\def\O{\mathcal{O}}
\def\vec{\mathbf}
\def\i{\mathrm{i}}
\def\e{\mathrm{e}}
\def\P{\mathcal{P}}

\def\sign{\rm sign}

\usepackage[errorstop]{feynmp}
\newcommand{\lsim}
{\;\raisebox{-.3em}{$\stackrel{\displaystyle <}{\sim}$}\;}

\begin{document}

\thispagestyle{empty}

\begin{flushright}
{
\small
KCL-PH-TH/2022-06
}
\end{flushright}

\vspace{0.4cm}

\begin{center}
\Large\bf\boldmath
Dissipation of oscillating scalar backgrounds in an FLRW universe
\unboldmath
\end{center}

\vspace{-0.2cm}

\begin{center}
{Zi-Liang Wang$^{*1}$\blfootnote{$^*$~ziliang.wang@just.edu.cn} 
and Wen-Yuan Ai$^{\dagger 2}$\blfootnote{$^\dagger$~wenyuan.ai@kcl.ac.uk, corresponding author} \\
\vskip0.4cm
{\it $^1$Department of Physics, School of Science,\\
Jiangsu University of Science and Technology,
Zhenjiang, 212003, China} \par
{\it $^2$Theoretical Particle Physics and Cosmology, King’s College London,\\ Strand, London WC2R 2LS, UK}\\
\vskip1.4cm}
\end{center}

\begin{abstract}
We study the dissipation of oscillating scalar backgrounds in a spatially flat Friedmann–Lema\^{i}tre–Robertson–Walker universe using non-equilibrium quantum field theory. To be concrete, a $Z_2$-symmetric two-scalar model with quartic interactions is used. For quasi-harmonic oscillations, we adopt the multi-scale analysis to obtain {\it analytical} approximate expressions for the evolution of the scalar background in terms of the retarded self-energy and retarded proper four-vertex function. Different from the case in flat spacetime, we find that in an expanding universe the condensate decay in this model can be complete only if the imaginary part of the retarded self-energy is not negligibly small. The microphysical interpretation of the imaginary parts of the retarded self-energy and retarded proper four-vertex function in terms of particle production is also discussed.

\end{abstract}

\newpage
\tableofcontents

\section{Introduction}

Scalar fields play prominent roles in both particle physics and cosmology. In the standard model (SM),  there is a known scalar field, the Higgs field, which is crucial for mass generation. Scalar fields are also assumed to be crucial in explaining a variety of phenomena beyond the SM. For example, in the Peccei-Quinn mechanism \cite{Peccei:1977hh}, which can elegantly solve the strong CP problem \cite{Peccei:2006as,Kim:2008hd}, a (pseudo)scalar field called axion \cite{Weinberg:1977ma,Wilczek:1977pj} is generally predicted.\footnote{It is argued recently in Ref.~\cite{Ai:2020ptm} that the strong CP problem may not exist at all.} In the inflationary scenario for the very early Universe \cite{Starobinsky:1980te,Guth:1980zm,Linde:1981mu,Albrecht:1982wi}, most of the inflation models assume that the inflaton is a scalar field~\cite{Martin:2013tda}. Further, gauge singlet scalars~\cite{McDonald:1993ex} could also account for part or all of the Dark Matter \cite{Hui:2016ltb}.

In the standard single-field slow-roll inflation, the inflaton field initially has a non-vanishing field displacement from its equilibrium value and slowly rolls down to the latter. Near the minimum, it oscillates and transfers its kinetic and potential energy to perturbative fluctuations of fields coupled to it, leading to a dramatic production of particles. The associated thermalization process reheats the Universe and the standard big-bang picture of the Universe follows. For large elongations, oscillations break adiabaticity, leading to the phenomenon of parametric resonance~\cite{Kofman:1994rk,Shtanov:1994ce,Boyanovsky:1994me,Boyanovsky:1996sq,Kofman:1997yn}. Particle production during this period is non-perturbative and the corresponding reheating process is usually called preheating.\footnote{A different mechanism for non-perturbative particle production is the tachyonic preheating due to the spinodal instability~\cite{Felder:2000hj,Felder:2001kt,Tomberg:2021bll,Koivunen:2022mem}.}  After preheating, the oscillation amplitude becomes small enough such that the process of particle production would finally become perturbative. In this work, we study the dissipation of oscillating scalar backgrounds in the latter regime of small elongations such that one can apply the small-field expansion in which the condensate-dependent masses of fluctuation fields are treated as perturbative terms compared to the condensate-independent masses.\footnote{For the dissipative behavior of a scalar condensate in the slow-roll phase, see Refs.~\cite{Morikawa:1985mf,Morikawa:1986rp,Gleiser:1993ea,Berera:2007qm,Buldgen:2019dus}.} 

Parametric resonance due to oscillating backgrounds has been studied thoroughly in the aforementioned classic papers on preheating. Generically, studies on non-perturbative particle production are usually based on the Bogoliubov method~\cite{Dolgov:1989us}, the functional Schr\"{o}dinger approach~\cite{Traschen:1990sw}, in particular the Floquet theory on analyzing equations of motion for mode functions. Perturbative particle production is usually studied by the time-dependent perturbation theory, see, e.g., Refs.~\cite{Ichikawa:2008ne,Aoki:2022dzd}. These methods typically assume a fixed classical background and the computation is usually done at zero temperature, thus missing the backreaction effects and thermal effects.  Reformulating the approaches in terms of particle distributions offers a way to incorporate thermal effects~\cite{Laine:2016hma,Emond:2018ybc,Moroi:2020bkq}. A natural framework with the above two effects being simultaneously taken into account does exist and is provided by the Closed-Time-Path (CTP) formalism~\cite{Schwinger:1960qe,Keldysh:1964ud}. Some earlier studies on dissipation due to particle production using the CTP formalism can be found in Refs.~\cite{Calzetta:1989vs,Paz:1990jg,Paz:1990sd,Boyanovsky:1994me}, while Refs.~\cite{Greiner:1996dx,Yokoyama:2004pf,Mukaida:2012qn} focus on in particular oscillating backgrounds. In principle, the non-equilibrium dynamics can be understood by solving the coupled equations of motion for the one-point function (of the scalar field that forms condensate) and for the two-point functions (the Kadanoff-Baym equations). However, due to the limited ability to solve these equations analytically, clear analytical relations between the particle production rates or the condensate evolution and the various microscopic quantities like self-energies and proper four-vertex functions, have never been rigorously derived.  

Important progress has been made recently in Ref.~\cite{Ai:2021gtg} where the authors were able to solve the equation of motion for the scalar condensate {\it analytically} in the small-field regime and when the oscillation is quasi-harmonic.\footnote{For a complementary understanding to Ref.~\cite{Ai:2021gtg}, see the recent work~\cite{Kainulainen:2021eki} where the authors study the condensate evolution beyond the small-field regime and solve the coupled equations of motion for the condensate and two-point functions numerically, capturing effects from both parametric resonance and spinodal instability.} A crucial development made in Ref.~\cite{Ai:2021gtg} is introducing the multi-scale analysis~\cite{Bender,Holmes} to solving the non-local equation of motion for the condensate. This method assumes that physical processes happen on different time scales. Specifically, it assumes that the oscillating amplitude and frequency change slightly during one single oscillation and the window provided by the kernels of the non-local terms in the condensate equation of motion. This is usually the case for the dissipating system under study when the coupling constants are perturbatively small. The obtained condensate evolution is expressed in terms of the retarded self-energy and retarded proper four-vertex function for the scalar that forms condensate.

In this paper we generalize the work~\cite{Ai:2021gtg} to a flat Friedmann–Lema\^{i}tre–Robertson–Walker (FLRW) universe. This generalization is of closer relevance to the perturbative reheating process in the early Universe. Although particle production via parametric resonance may be more efficient than the perturbative particle production, the latter still determines whether the dissipation of inflaton is complete or not as well as some initial conditions right after reheating. On the other hand, the scalar field under study is not necessarily the inflaton, but can be other fields whose oscillations may be relevant for the production of Dark Matter. Therefore, this work also serves as a first-principle, and perhaps also more rigorous than alternative methods, framework for studying perturbative production of Dark Matter from an oscillating scalar field~\cite{Almeida:2018oid,Garcia:2020eof,Moroi:2020has,Mambrini:2021zpp,Ahmed:2021fvt}.\footnote{Examinations on this scenario focusing on non-perturbative production of Dark Matter are given in Refs.~\cite{Lebedev:2021zdh,Lebedev:2021tas}.} The outline of the paper is as follows. In the next section, we introduce our model and review the derivation of the condensate equation of motion using the two-particle-irreducible (2PI) effective action~\cite{Cornwall:1974vz,Ai:2021gtg}. In Sec.~\ref{sec:solveEoM}, we solve the condensate equation of motion for a static universe and a radiation-dominated universe using the multi-scale analysis with the time-dependence in the microscopic quantities neglected. We then discuss the cosmological meanings of the obtained solutions. Since the microscopic quantities depend on the temperature, neglecting their time-dependence is therefore not self-consistent if the temperature is evolving. To fully take into account the time-dependence, one needs to know the closed expressions of all the microscopic quantities. In Sec.~\ref{app1},  we study the effects from the time-dependence of the microscopic quantities in a simple situation.  Finally, we present our conclusions in Sec.~\ref{sec:Conc}. For completeness, we also include a discussion of the solutions for a matter-dominated universe in the Appendix.

\section{Model and the condensate equation of motion}

\label{sec:model and EoM}

We consider the following action
\begin{align}
S[\Phi, \chi]=\int \d^{4}x \,\sqrt{-g} \biggl[
	\frac{1}{2} (\partial_\mu \Phi) (\partial^\mu \Phi)
	+ \frac{1}{2} (\partial_\mu \chi) (\partial^\mu \chi)
	-V(\Phi,\chi)
	\biggr] \, ,
\label{action}
\end{align}
where
\begin{equation}
V(\Phi,\chi)= \frac{m_\phi^2}{2} \Phi^2
	+ \frac{m_\chi^2}{2} \chi^2
	+ \frac{\lambda_\phi}{4!} \Phi^4
	+ \frac{\lambda_\chi}{4!} \chi^4
	+ \frac{g}{4} \Phi^2 \chi^2\,.
\end{equation}
Here $\Phi$ and $\chi$ are two real scalars. This model is frequently considered in phenomenological studies as one of the Higgs portal models~\cite{Lebedev:2021xey}.  For simplicity, we do not include the non-minimal coupling with gravity.
For a spatially flat FLRW universe
\begin{equation}
\d{s}^2 = \d{t}^2 - a^2(t) \d{\vec x}^2\,,
\end{equation}
and $\sqrt{-g}=a^3(t)$. In this paper, we take the spacetime background as fixed. This means that there are other components other than the oscillating scalar field that dominate the energy density and thus determine the universe expansion. In the case that the oscillating scalar is the inflaton, this can be the case at the very late stage of reheating when the radiation produced from the earlier stage of the reheating, e.g., preheating, dominates the energy density.

The scalar field $\Phi$ is assumed to possess a non-vanishing expectation value $\langle{\Phi}\rangle \equiv \varphi$ and will be called the inflaton.  Expanding $\Phi = \varphi + \phi$, we can view $\varphi$ as a background and $\phi$, $\chi$ as fluctuations about this background. Particles are then defined as excitations of the fluctuation fields $\phi$ and $\chi$. The total action can be written as $S[\Phi,\chi]=S_\varphi[\varphi]+S_{\phi\chi}[\phi,\chi;\varphi]$ where
\begin{subequations}
\begin{align}
&S_\varphi[\varphi]  =\int\d^4 x\, a^3(t) \left[\frac{1}{2}(\partial_\mu\varphi)(\partial^\mu\varphi)-\frac{1}{2}m^2_\phi\varphi^2-\frac{\lambda_\phi}{4!}\varphi^4\right]\,, \\
&S_{\phi\chi}[\phi,\chi;\varphi]  =\int\d^4 x\,a^3(t)\left[\frac{1}{2}(\partial_\mu\phi)(\partial^\mu\phi)+\frac{1}{2}(\partial_\mu\chi)(\partial^\mu\chi)-V_{\phi\chi}(\phi,\chi)\right]\,, \label{eq:Sphichi}
\end{align}
\end{subequations}
with
\begin{align}\
\label{eq:Vhphi}
V_{\phi\chi}(\phi,\chi) &=
\frac{1}{2} \left(m_\phi^2 + \frac{\lambda_{\phi}}{2}\varphi^2\right) \phi^2+\frac{1}{2} \left(m^2_\chi +  \frac{g}{2} \varphi^2\right) \chi^2+\frac{\lambda_\phi}{3!}\varphi\phi^3+\frac{g}{2}\varphi\phi\chi^2
+ \frac{\lambda_\phi}{4!} \phi^4
+ \frac{\lambda_\chi}{4!} \chi^4\notag\\
&\quad +\frac{g}{4}\phi^2\chi^2+({\rm linear\ terms\ in\ fluctuations})\,.
\end{align}
The linear terms in fluctuations would not contribute to the perturbative diagrammatic expansion of the effective action.\footnote{For a clear explanation on this point in the case of 1PI effective action, see, e.g.,  Ref.~\cite{Peskin:1995ev}.} 
The oscillation of $\varphi$ could induce particle production for both $\phi$ and $\chi$, either through the time-dependent mass terms or the interacting terms $\varphi\phi^3$, $\varphi\phi\chi^2$. The small-field regime we consider in this work is defined by the requirement that the $\varphi$-dependent mass term is much smaller than the  $\varphi$-independent mass term of the same particle type. (At finite temperature, one may consider thermal corrections to the $\varphi$-independent mass terms.)
Therefore, we can take the $\varphi$-dependent mass terms as perturbations such that in the perturbative expansion of the effective action they are on an equal footing with the other interacting terms. In particular, this means that one can expand the two-point functions in the background field $\varphi$ and at the leading order they are independent of $\varphi$. Below we briefly review the derivation of the equation of motion for $\varphi$ using the CTP formalism~\cite{Ai:2021gtg}. A reader not concerned with this derivation may take Eq.~\eqref{eq:condensate eom2}
as a starting point.

The CTP formalism has been widely applied to the studies of baryogenesis, especially in the form of leptogenesis (see, e.g., Refs.~\cite{Buchmuller:2000nd,Prokopec:2003pj,Prokopec:2004ic,DeSimone:2007gkc,Garny:2009rv,Garny:2009qn,Anisimov:2010aq,Beneke:2010wd,Beneke:2010dz,Anisimov:2010dk}).\footnote{Recently, it has been applied to study processes in stellar media~\cite{Chadha-Day:2022inf}.} For reviews, see Refs.~\cite{Chou:1984es,Calzetta:1986cq,Berges:2004yj}. In the conventional zero-temperature quantum field theory, there are known asymptotic states at far past and far future, and most physical problems are about the transition amplitudes between the asymptotic states. They belong to the boundary-value problem. In non-equilibrium quantum field theory, the initial states are prepared and the final states are unknown {\it a priori} and can only be determined by the evolution itself. Thus the non-equilibrium dynamics is an initial-value problem. This is the reason why a closed-time path needs be introduced in non-equilibrium quantum field theory.

\begin{figure}[ht]
    \centering
    \includegraphics[scale=0.3]{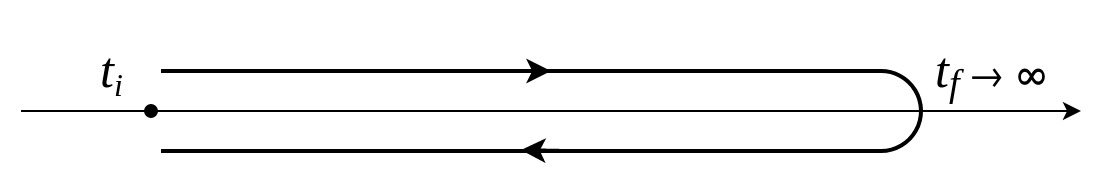}
    \caption{The Keldysh contour $\mathcal{C}$ for the generating functional in the CTP formalism.}
    \label{fig:keldyshcontour}
\end{figure}

The initial value is usually given by a density matrix at a given time $\rho_D(t_i)$ in a mixed (${\rm Tr}\{\rho^2_D(t_i)\}<1$) or pure (${\rm Tr}\{\rho^2_D(t_i)\}=1$) state. In the Heisenberg picture, operators evolve with time while the states do not. The expectation value of an observable $\mathcal{O}$ at time $t$ is given by
\begin{align}
\label{eq:expetation}
    \langle \mathcal{O}(t)\rangle ={\rm Tr} \left\{\rho_D(t_i)\mathcal{O}(t)\right\}\,,
\end{align}
where $\mathcal{O}(t)=\exp(\i H(t-t_i))\mathcal{O}(t_i)\exp(-\i H(t-t_i))$. These expectation values can be obtained by a generating functional formulated on a closed time contour $\mathcal{C}$, as illustrated in Fig.~\ref{fig:keldyshcontour}. The full information in a quantum field system is encoded in all its correlation functions. For most purposes, it is sufficient to study the one-point function, $\varphi$, and the connected two-point functions, $\Delta_\phi(x,y)\equiv\langle \Phi(x)\Phi(y)\rangle_{\rm c}$ and $\Delta_\chi(x,y)\equiv \langle\chi(x)\chi(y)\rangle_{\rm  c}$. Therefore the non-equilibrium dynamics is usually given by the coupled equations of motion for the one- and two-point functions. The equations of motion for these quantities are determined by the 2PI effective action, $\Gamma_{\rm 2PI}[\varphi,\Delta_\phi,\Delta_\chi]$~\cite{Cornwall:1974vz,Berges:2004yj},\footnote{Remember in the equation of motion for $\varphi$, one has to take the limit $\varphi^+(x)=\varphi^-(x)$ where $\pm$ indicate the forward and backward branches of the Keldysh contour, respectively~\cite{Calzetta:1986cq}.}
\begin{subequations}
\label{eq:eom}
\begin{align}
\frac{\delta\Gamma_{\rm 2PI}[\varphi,\Delta_\phi,\Delta_\chi]}{\delta\varphi(x)}&=0\,,\\
\frac{\delta\Gamma_{\rm 2PI}[\varphi,\Delta_\phi,\Delta_\chi]}{\delta \Delta_\phi(x,y)}&=0\,,\\
\frac{\delta\Gamma_{\rm 2PI}[\varphi,\Delta_\phi,\Delta_\chi]}{\delta \Delta_\chi(x,y)}&=0\,.
\end{align}
\end{subequations}
Solving on-shell for the two-point functions as functionals of the condensate
\begin{equation}
\frac{ \delta \Gamma_\text{2PI} }{ \delta \Delta_\phi } \biggr|_{\Delta_\phi = \Delta_\phi[\varphi]}=0 \, ,
\qquad\quad
\frac{ \delta \Gamma_\text{2PI} }{ \delta \Delta_\chi } \biggr|_{\Delta_\chi = \Delta_\chi[\varphi]}=0 \, ,
\label{symbolic 2pt eom}
\end{equation}
and plugging them back into the 2PI effective action gives the equation of motion for the condensate
\begin{equation}
\label{eq:eom-general}
\left.\frac{ \delta \Gamma_\text{2PI}[\varphi,\Delta_\phi,\Delta_\chi] }{ \delta \varphi(x) }\right|_{\Delta_\phi=\Delta_\phi[\varphi],\,\Delta_\chi=\Delta_\chi[\varphi]}= 0 \, .
\end{equation}
Note that we have assumed that all external sources are vanishing. Otherwise the above equations of motion would have nonvanishing terms on the RHS related to the sources~\cite{Garbrecht:2015cla,Millington:2019nkw}. In some discussions for dissipation, one would have to introduce nonvanishing sources to switch on perturbations at the initial time, see e.g., Ref.~\cite{Calzetta:1986ey}. The 1PI or 2PI effective action can also be used to study radiative corrections to false vacuum decay~\cite{Garbrecht:2015oea,Garbrecht:2015yza,Ai:2018guc,Ai:2020sru,Cruz:2022ext}.

When the oscillation amplitudes are small, one can perform a perturbative expansion in $\varphi$ in Eqs.~\eqref{symbolic 2pt eom} and at the leading order, the connected two-point functions are independent of $\varphi$. Substituting the leading-order connected two-point functions into Eq.~\eqref{eq:eom-general}, one obtains an equation of motion for the condensate in which propagators are the free thermal ones (see Eqs.~\eqref{eq:free-thermal-propagators}). Properly truncating the 2PI effective action for $\varphi$, the condensate equation of motion in flat spacetime is given as~\cite{Ai:2021gtg} \begin{equation}
\ddot{\varphi}(t) + M_\phi^2 \varphi(t)
    + \frac{\lambda_\phi \varphi^3(t) }{6}
	+ \int_{t_i}^{t}
	    \d t' \, \pi_{\rm R}(t-t') \varphi(t')
	+ \frac{\varphi(t) }{6}
	\int_{t_i}^{t} \d t' \,
	    v_{\rm R}(t-t') \varphi^2(t') = 0 \, ,
\label{eq:condensate eom}
\end{equation}
where~$t_i$ is the initial time at which initial conditions~$\varphi(t_i)$ and~$\dot{\varphi}(t_i)$
are specified, $M_\phi^2$ is the thermal-corrected mass for the $\Phi$ scalar, $\pi_{\rm R}$ is the retarded {\it self-energy} and, $v_{\rm R}$ is the retarded {\it proper four-vertex function}. These quantities will be defined specifically below. As we shall see in the next section, the last two quantities play distinguished roles in the inflaton decay. Note that since we have performed the small-field expansion for the two-point functions and taken the leading order results, the propagators in the $2{\rm PI}$ effective action are independent of $\varphi$ and therefore the background field only appears in the vertices but not in the propagators. As a result, in Eq.~\eqref{eq:condensate eom} we do not have the usual effective potential~\cite{Coleman:1973jx,Quiros:1999jp}. For example, the familiar one-loop effective potential would be obtained when the propagators have a dependence on the background field that is approximated as constant.

We shall assume that the thermal bath is large such that it is in equilibrium all the time. Diagrammatically, the two-point functions at the leading level of the small-field expansion and the two-loop level of the 2PI effective action are simply given by the free thermal equilibrium propagators (the Schwinger-Keldysh polarity indices $A,B$ take values of $+,-$)
\begin{equation}
\begin{tikzpicture}[baseline={0cm-0.5*height("$=$")}]
\draw[thick] (-0.6,0) -- (0.6,0) ;
\filldraw (-0.6,0) circle (1.5pt) node[below] {$\scriptstyle(x,A)$} ;
\filldraw (0.6,0) circle (1.5pt) node[below] {$\scriptstyle(x',B)$} ;
\end{tikzpicture}
\equiv D_\phi^{AB}(x-x') \, ,
\qquad \qquad
\begin{tikzpicture}[baseline={0cm-0.5*height("$=$")}]
\draw[thick,densely dashed] (-0.6,0) -- (0.6,0) ;
\filldraw (-0.6,0) circle (1.5pt) node[below] {$\scriptstyle(x,A)$} ;
\filldraw (0.6,0) circle (1.5pt) node[below] {$\scriptstyle(x',B)$} ;
\end{tikzpicture}
\equiv
D_\chi^{AB}(x-x') \, ,
\label{thermal propagators}
\end{equation}
where
\begin{subequations}
\label{eq:free-thermal-propagators}
\begin{align}
D_{\phi,\chi}^{++}(x-x')
=&\int\frac{\d^4 k}{(2\pi)^4}\,
\e^{-\i k(x-x')}
\left[\frac{\i}{k^2 -M_{\phi,\chi}^2+\i\varepsilon}+2\pi f_{\rm B}(|k_0|)\delta\left( k^2 - M_{\phi,\chi}^2 \right) \right] \, ,\label{D++ thermal}\\
D_{\phi,\chi}^{+-}(x- x')
=&\int\frac{\d^4 k}{(2\pi)^4 } \,
\e^{-\i k (x-x')}\left[2\pi f_{\rm B}(k_0) \sign(k_0) \,\delta\left(k^2 - M_{\phi,\chi}^2 \right) \right] \, ,\label{D+- thermal}
\end{align}
\end{subequations}
with the remaining components being complex
conjugates,~$D^{--}_{\phi,\chi}(x-x') = \left[ D^{++}_{\phi,\chi}(x-x') \right]^*$, $D^{ -+}_{\phi,\chi}(x-x') = \left[ D^{+-}_{\phi,\chi}(x-x') \right]^*$.
Here~$f_{\rm B}(\omega) = 1/\left(\e^{\omega/T} - 1 \right)$
is the Bose-Einstein distribution
and the thermal
masses are
\begin{subequations}
\begin{align}
M_\phi^2 &=m_\phi^2-\frac{\i \lambda_\phi}{2}
\begin{tikzpicture}[baseline={0cm-0.5*height("$=$")}]
\draw[thick] (0,0) circle (0.5) ;
\filldraw (0.5,0) circle (1.5pt) node[right] {$\scriptstyle (x,+)$} ;
\end{tikzpicture}
-\frac{\i g}{2}
\begin{tikzpicture}[baseline={0cm-0.5*height("$=$")}]
\draw[thick, dashed] (0,0) circle (0.5) ;
\filldraw (0.5,0) circle (1.5pt) node[right] {$\scriptstyle (x,+)$} ;
\end{tikzpicture}
=m_\phi^2 +\frac{ (\lambda_\phi \!+\! g) T^2}{24}
\, ,\label{thermal Mphi}\\
M_\chi^2 &=m_\chi^2-\frac{\i \lambda_\chi}{2}
\begin{tikzpicture}[baseline={0cm-0.5*height("$=$")}]
\draw[thick,dashed] (0,0) circle (0.5) ;
\filldraw (0.5,0) circle (1.5pt) node[right] {$\scriptstyle (x,+)$} ;
\end{tikzpicture}
-\frac{
\i g}{2}
\begin{tikzpicture}[baseline={0cm-0.5*height("$=$")}]
\draw[thick] (0,0) circle (0.5) ;
\filldraw (0.5,0) circle (1.5pt) node[right] {$\scriptstyle (x,+)$} ;
\end{tikzpicture}
=m_\chi^2 +\frac{ (\lambda_\chi \!+\! g) T^2}{24}
\, ,
\label{thermal Mchi}
\end{align}
\end{subequations}
where the last step in each equation is only satisfied in the high-temperature limit. The various self-energies and proper four-vertex functions read
\begin{subequations}
\begin{align}
\label{PiDef}
\Pi_{AB}(x-x')
	&=
\
-
\,
\frac{\i (AB)\lambda_\phi^2}{6}
\,
\begin{tikzpicture}[baseline={0cm-0.5*height("$=$")}]
\draw[thick] (0,0) circle (0.5) ;
\draw[thick] (-0.5,0) -- (0.5,0) ;
\filldraw (-0.5,0) circle (1.5pt) node[left] {$\scriptstyle(x,A)$} ;
\filldraw (0.5,0) circle (1.5pt) node[right] {$\scriptstyle(x',B)$} ;
\end{tikzpicture}
\
-
\
\frac{\i (AB)g^2}{2}
\,
\begin{tikzpicture}[baseline={0cm-0.5*height("$=$")}]
\draw[thick,dashed] (0,0) circle (0.5) ;
\draw[thick] (-0.5,0) -- (0.5,0) ;
\filldraw[fill=black] (0.5,0) circle (1.5pt) node[right] {$\scriptstyle(x',B)$} ;
\filldraw[fill=black] (-0.5,0) circle (1.5pt) node[left] {$\scriptstyle(x,A)$} ;
\end{tikzpicture}
\,,\\ 
\label{VDef}
V_{AB}(x-x')
&=
\
-
\
\frac{3\i (AB) \lambda_\phi^2}{2}
\,
\begin{tikzpicture}[baseline={0cm-0.5*height("$=$")}]
\draw[thick] (0,0) circle (0.5) ;
\filldraw[fill=black] (0.5,0) circle (1.5pt) node[right] {$\scriptstyle(x',B)$} ;
\filldraw[fill=black] (-0.5,0) circle (1.5pt) node[left] {$\scriptstyle(x,A)$} ;
\end{tikzpicture}
\
-
\
\frac{3\i (AB) g^2}{2}
\,
\begin{tikzpicture}[baseline={0cm-0.5*height("$=$")}]
\draw[thick, dashed] (0,0) circle (0.5) ;
\filldraw[fill=black] (0.5,0) circle (1.5pt) node[right] {$\scriptstyle(x'\!,B)$} ;
\filldraw[fill=black] (-0.5,0) circle (1.5pt) node[left] {$\scriptstyle(x,A)$} ;
\end{tikzpicture}
\,.
\end{align}
\end{subequations}
The retarded self-energy and retarded proper four-vertex function are then defined as
\begin{subequations}\label{piR+vR def}
\begin{align}
&
\Pi_{\rm R}(x-x')
    = \Pi_{++}(x-x') + \Pi_{+-}(x-x')
	= \theta(t-t')
	    \left[ - \Pi_{-+}(x-x')
	        + \Pi_{+-}(x-x')  \right] \, ,
\label{piR def}
\\
&
V_{\rm R}(x-x')
    = V_{++}(x-x') + V_{+-}(x-x')
	= \theta(t-t')
	    \left[ - V_{-+}(x-x')
	        + V_{+-}(x-x')  \right] \, .
\label{vR def}
\end{align}
\end{subequations}
Since the condensate is homogeneous, we have the spatial translation symmetry. We therefore define
\begin{equation}
\pi_{\rm R}(t-t')
	= \int \d^{3}x \, \Pi_{\rm R}(x-x') \, ,
\qquad
v_{\rm R}(t-t')
	= \int \d^{3}x \, V_{\rm R}(x-x') \, .
\end{equation}
As we shall see, it is the Fourier transforms of the retarded self-energy and four-vertex function evaluated at specific frequencies, defined as
\begin{equation}
\label{PiTildeDef}
\widetilde{\pi}_{\rm R}(\omega) = \int_{-\infty}^{\infty} \
	\d t'\, \e^{\i\omega(t-t')} \pi_{\rm R}(t-t') \, ,
\qquad 
\widetilde{v}_{\rm R}(\omega) = \int_{-\infty}^{\infty} \d t' \, \e^{\i\omega(t-t')} v_{\rm R}(t-t') \, ,
\end{equation} that will appear in the condensate evolution, see Eqs.~\eqref{eq:FourierTransf} and~\eqref{eq:important_quantities} below. 

It is usually difficult to obtain closed-form expressions for these Fourier transforms. For the model we consider, some estimates and discussions on them can be found in, e.g., Refs.~\cite{Boyanovsky:2004dj,Drewes:2013bfa,Parwani:1991gq,Drewes:2013iaa}. The imaginary part of the proper four-vertex is known exactly~\cite{Boyanovsky:2004dj},
\begin{align}
   {\rm Im}\left[ \widetilde{v}_{\rm R}(\omega) \right] =&-\theta\left(\omega^2 - 4 M_\phi^2 \right)
	\frac{ 3 \lambda_\phi^2 }{ 32\pi }
		\sqrt{ 1 -\frac{4 M_\phi^2}{\omega^2} }\,
		\left[ 1 + 2f_{\rm B} \left( \frac{\omega}{2} \right) \right] 
\nonumber \\
&   \hspace{1.5cm}
	- \theta\left(\omega^2 - 4
	    M_\chi^2 \right)
		\frac{3 g^2 }{ 32\pi }
		\sqrt{ 1 -\frac{4 M_\chi^2}{\omega^2} }
		\,\left[ 1 + 2f_{\rm B} \left( \frac{\omega}{2} \right) \right] \, .
\label{Im vR omega T}
\end{align}
In the limit of $T=0$, 
\begin{equation}
\label{eq:Im-v-T0}
  {\rm Im}\left[ \widetilde{v}_{\rm R}(\omega) \right] =	- \theta\bigl(\omega^2 - 4 m_\phi^2 \bigr)
	\frac{ 3 \lambda_\phi^2 }{ 32\pi }
		\sqrt{ 1 - \frac{4 m_\phi^2}{\omega^2} }
	- \theta\bigl(\omega^2 \!-\! 4
	m_\chi^2 \bigr)
		\frac{ 3 g^2 }{ 32\pi }
		\sqrt{ 1 - \frac{4 m_\chi^2}{\omega^2} }.
\end{equation}
The real part of the proper four-vertex has been studied in, e.g., Ref.~\cite{Drewes:2013bfa} and can be calculated from the imaginary part via the Kramers–Kronig relation~\cite{Bellac:2011kqa}
\begin{align}
    {\rm Re} \left[\widetilde{v}_{\rm R}(\omega)\right]=\P \int\frac{\d \omega'}{\pi}\,\frac{{\rm Im}\left[\widetilde{v}_{\rm R}(\omega')\right]}{\omega'-\omega}\,,
\end{align}
where $\P$ denotes the Cauchy principal value. The self-energy corresponding to the ``setting sun'' diagrams is much more complicated.  Closed-form expressions for the self-energy $\widetilde{\pi}_{\rm R}(\omega)$, to our best of knowledge, are not known in the literature. In a work that is still in progress, we are trying to obtain closed-form expressions for the imaginary and real parts of the self-energy. We have obtained a closed-form expression for ${\rm Im}[\widetilde{\pi}_{\rm R}(\omega)]$ with only the contribution from the first diagram in Eq.~\eqref{PiDef} when $\omega=M_\phi$ (this is the only relevant case, see Eq.~\eqref{eq:important_quantities})~\cite{AL}, 
\begin{align}
\label{eq:Im-pi}
    {\rm Im} \left[\widetilde{\pi}_{\rm R,\phi}(M_\phi)\right]=-\frac{\lambda_\phi^2 T^2}{128\pi^3}{\rm Li}_2\left(\e^{-M_\phi/T}\right)\,,
\end{align}
where ${\rm Li}_2 (z)$ is the dilogarithm function defined as
\begin{align}
    {\rm Li}_2(z) =-\int ^{1} _{0} \frac{\d x}{x} \ln (1-z\,x)\,,
\end{align}
for $z\in (-\infty, 1)$. The dilogarithm function has the following asymptotic behaviors
\begin{subequations}
\begin{align}
       \label{eq:Li2-highT}
     &\text{Li}_2(\e ^{- M_{\phi}/T}) = \frac{\pi ^2}{6} + O\left[\frac{M_{\phi}}{T}\ln(\frac{M_{\phi}}{T})\right]\quad {\rm for\ }T\gg M_\phi\,,\\
     \label{eq:Li2-lowT}
     &\text{Li}_2(\e ^{- M_{\phi}/T}) = \e ^{- M_{\phi}/T} + O\left(\e ^{-2 M_{\phi}/T}\right)\quad {\rm for\ } T\ll M_\phi\,.
\end{align}
\end{subequations}

The high-temperature limit of Eq.~\eqref{eq:Im-pi} is consistent with the result given in Refs.~\cite{Parwani:1991gq,Drewes:2013iaa}. For $T\rightarrow 0$ the imaginary part of the self-energy evaluated at $\omega=M_\phi$ vanishes. This is reasonable because the imaginary part has the optical-theorem interpretation~\cite{Peskin:1995ev} and at zero temperature, the particles in the loop in the diagrams of Eq.~\eqref{PiDef} (more correctly, of Eq.~\eqref{eq:diag-self-energy} below) cannot be on-shell when the external particle has four-momentum $(M_\phi,\vec{0})$. For more discussions on the microscopic interpretation of the condensate dissipation, see Sec.~\ref{sec:micro-inter}.

The quantities in Eq.~\eqref{eq:important_quantities} depend on the plasma temperature.  In this work, to avoid applying uncertain values of the quantities in Eq.~\eqref{eq:important_quantities}, we shall simply take them as input parameters when looking into the behaviors of the condensate evolution. As a consequence, the effects from the time-dependence of the microscopic quantities are neglected. These effects could be important in reheating.\footnote{In the frequently-used treatment of perturbative reheating, one usually assumes a simple Markovian equation for the condensate, $\ddot{\varphi}+3H\dot{\varphi}+V'(\varphi)+\Gamma_\phi\dot{\varphi}=0$ where $\Gamma_\phi$ is assumed to be the one-body decay rate in vacuum for $\Phi$~\cite{Chung:1998rq,Giudice:2000ex} which is constant. An improvement is to take $\Gamma_\varphi$ as the vacuum decay rate in presence of a fixed oscillating background~\cite{Ichikawa:2008ne}. This slightly improved treatment has been used by many authors (see, e.g., Refs~\cite{Nurmi:2015ema,Almeida:2018oid,Garcia:2020wiy,Aoki:2022dzd,Lebedev:2022ljz}). Still, in this improved calculation, $\Gamma_\varphi$ depends on time only through its dependence on $\varphi$ and the time-dependence of $\Gamma_\varphi$ through the evolving temperature is still neglected. To take into account the latter, some authors formally assume a parameterization of $\Gamma_\varphi$ in powers of $T$~\cite{Drewes:2014pfa,Co:2020xaf,Ming:2021eut,Barman:2022tzk}.} But taking into account these effects is only possible when the exact dependence on $T$ in all the Fourier transforms in Eq.~\eqref{eq:important_quantities} are known. In Sec.~\ref{app1}, we study this time-dependence of the microscopic quantities in a very simple situation. A more dedicate study of perturbative reheating applying our theoretical methods developed here is left for future work~\cite{AL}.

In an expanding universe, one in principle needs to rederive the condensate equation of motion following Ref.~\cite{Ai:2021gtg}. However, this would make the problem too complicated. For example, because of the time-dependent metric, the free thermal equilibrium propagators would take a time-translation-variant form, depending on the specific form of $a(t)$. Below, we shall assume that the microscopic time scales are much smaller than the Hubble time so that, one can assume a flat spacetime background in the non-local terms in Eq.~\eqref{eq:condensate eom} but only take into account the effects from the expanding universe in the classical part. Therefore, we will use the following equation of motion for the condensate $\varphi$, 
\begin{equation}
\ddot{\varphi}(t) +  M_\phi^2 \varphi(t) + 3H\dot{\varphi}(t)
    + \frac{\lambda_\phi \varphi^3(t) }{6}
	+ \int_{t_i}^{t}
	    \d t' \, \pi_{\rm R}(t-t') \varphi(t')
	+ \frac{\varphi(t) }{6}
	\int_{t_i}^{t} \d t' \,
	    v_{\rm R}(t-t') \varphi^2(t') = 0 \, .
\label{eq:condensate eom2}
\end{equation}
We shall note that all the masses and couplings in the above equation are assumed to be the renormalized ones. In the next section, we solve Eq.~\eqref{eq:condensate eom2} using the multi-scale analysis~\cite{Ai:2021gtg,Bender}.

\section{Solving the condensate equation of motion}

\label{sec:solveEoM}
Following~\cite{Ai:2021gtg}, we assume that the last three terms in Eq.~\eqref{eq:condensate eom2} are small compared with the mass term and the $\ddot{\varphi}(t)$ term. We also assume that the Hubble friction term proportional to the Hubble constant is small compared with the first two terms. In a theoretical viewpoint, this assumption sets the range of validity of our results. However, for realistic applications in the perturbative reheating process after preheating, the Hubble friction term can be shown to be small as follows. Typically, after preheating the Universe is filled with a high-temperature plasma and can be well assumed to be radiation dominated. Therefore, the Hubble constant is
\begin{align}
\label{eq:H of T}
    H^2\approx \frac{8\pi}{3M_{\rm pl}^2}\frac{g_{*}\pi^2 }{30} T^4\,,
\end{align}
where $M_{\rm pl}$ is the Planck mass and $g_*$ is the number of relativistic degrees of freedom. $g_*$ is model-dependent but to have an estimate we can take it to be $\O(100)$. The first and second terms in the equation of motion are $\sim M^2_\phi ||\varphi||$ where $||\varphi||$ denotes the oscillation amplitude, while the Hubble friction term is
\begin{align}
    H\dot{\varphi}\sim \frac{T^2}{M_{\rm pl} M_\phi}M_\phi^2||\varphi||\,.
\end{align}
Therefore if $T\ll\sqrt{M_{\rm pl}M_\phi}$, the Hubble friction term is suppressed by a factor of $T^2/(M_{\rm pl}M_\phi)$ compared with the first and second terms. A rough estimate on the reheating temperature based on instantaneous reheating~\cite{Kolb:1990vq} gives $T_{\rm RH}\sim \sqrt{\Gamma_\phi M_{\rm pl}}$ where $\Gamma_\phi$ is the decay rate of the inflaton and is proportional to powers of the couplings. There are a large parameter space (with $M_\phi$ large while the couplings small enough) in which $T\ll \sqrt{M_{\rm pl}M_\phi}$ can be satisfied.   

Now we solve Eq.~\eqref{eq:condensate eom2} step by step. First, for bookkeeping purposes, all small terms will be multiplied by a parameter $\varepsilon$. The equation of motion for the condensate $\varphi$ then reads
\begin{align}
  \ddot{\varphi}(t) +  M_\phi^2 \varphi(t) + 3 \varepsilon H\dot{\varphi}(t)
    + \varepsilon \frac{\lambda_\phi \varphi^3(t) }{6}
	+ \varepsilon \int_{t_i}^{t}
	    \d t' \, \pi_{\rm R}(t-t') \varphi(t')+ \varepsilon \frac{\varphi(t) }{6}
	\int_{t_i}^{t} \d t' \,
	    v_{\rm R}(t-t') \varphi^2(t') = 0 \, .
\label{eq:condensate eom3}  
\end{align}
Further, we expect that there is a hierarchy for the time scales in the   evolution of the condensate. The shorter time scale corresponds to the oscillation frequency $1/M_\phi$, and the longer time scale corresponds to the damping time scale. Therefore, we introduce two time variables, $t$ and $\tau =\varepsilon t$~\cite{Ai:2021gtg}.  Introducing a slow time variable $\tau$ allows us to correctly arrange terms in a perturbative expansion when there are terms with time derivatives. Once the perturbative calculation is done at a particular order, we will take $\varepsilon=1$ finally and then $\tau$ becomes the common physical time. The solution for the condensate $\varphi$ then  takes the following form
\begin{equation}
    \varphi(t) = \varphi(t,\tau;\varepsilon)\,.
\end{equation}
With this assumption, we are able to organize \eqref{eq:condensate eom3}  in powers of $\varepsilon$ as
\begin{align}
\frac{\partial ^2 \varphi(t,\tau;\varepsilon) }{\partial t^2} &+2 \varepsilon \frac{\partial ^2 \varphi(t,\tau;\varepsilon) }{\partial t \partial \tau}+ \varepsilon ^2 \frac{\partial^2 \varphi(t,\tau;\varepsilon)}{\partial \tau ^2} + M_\phi^2 \varphi(t,\tau;\varepsilon) \notag \\
& +3 \varepsilon H(\tau) \frac{\partial \varphi(t,\tau;\varepsilon) }{\partial t} + 3 \varepsilon ^2 H(\tau) \frac{\partial \varphi(t,\tau;\varepsilon)}{\partial \tau}+ \varepsilon \frac{\lambda_\phi }{6} \varphi^3(t,\tau;\varepsilon) \notag \\
& +\varepsilon \int_{t_i}^{t}
	    \d t' \, \pi_{\rm R}(t-t') \varphi(t',\tau ';\varepsilon)
	+ \varepsilon \frac{\varphi(t,\tau;\varepsilon) }{6}
	\int_{t_i}^{t} \d t' \,
	    v_{\rm R}(t-t')\varphi ^2(t',\tau ';\varepsilon) = 0 \,,
	    \label{eq:condensate eom4}
\end{align} 
where we have assumed that the Hubble parameter depends only on the slow time, i.e., 
\begin{equation}
    H=H(\tau)\,.
\end{equation}
This does not necessarily mean that the Hubble time scale is the same as the damping time scale.  It can be the case where the former is longer than latter.

Second, one can assume a power series solution of the form
\begin{equation}
    \varphi(t,\tau;\varepsilon)=\varphi_0 (t,\tau)+ \varepsilon \varphi_1 (t,\tau) + \varepsilon ^2 \varphi_2 (t,\tau)+...\,,
    \label{eq:varphi in powers of varepsilon}
\end{equation}
where 
\begin{equation}
    \varphi_n (t,\tau) =\frac{1}{n!}\frac{\partial ^n \varphi (t,\tau;\varepsilon)}{\partial \varepsilon ^n}\bigg|_{\varepsilon=0}
\end{equation}
are the coefficients. We aim to obtain the leading-order solution $\varphi_0$.

Third, to simplify the last two terms on the LHS of Eq.~\eqref{eq:condensate eom4}, we assume that the solution for the condensate $\varphi$ varies only slightly during the window provided by the kernel of the non-local terms. With this assumption, in the integrand of the non-local terms, we can Taylor-expand  $\varphi (t',\tau';\varepsilon)$  at $\tau$ as~\cite{Ai:2021gtg} 
\begin{align}
    \varphi (t',\tau';\varepsilon) 
    =\varphi (t',\tau;\varepsilon) +\varepsilon(t'-t)\frac{\partial \varphi (t',\tau;\varepsilon) }{\partial \tau} + \frac{\varepsilon^2}{2}(t'-t)^2 \frac{\partial^2 \varphi (t',\tau;\varepsilon) }{\partial \tau^2} +...\,,
\end{align}
where we have used the equality $\tau^{(\prime)} =\varepsilon t^{(\prime)}$. Then, up to the second order in $\varepsilon$, the non-local terms on the LHS of Eq.~\eqref{eq:condensate eom4} could be written as
\begin{subequations}
\begin{align}
    \varepsilon \int_{t_i}^{t}& \d t' \, \pi_{\rm R}(t-t') \varphi(t',\tau ';\varepsilon) \notag \\
    &=\varepsilon \int_{t_i}^{t} \d t' \, \pi_{\rm R}(t-t') \varphi(t',\tau ;\varepsilon)-\varepsilon ^2 \int_{t_i}^{t} \d t' \,(t-t') \pi_{\rm R}(t-t') \frac{\partial \varphi(t',\tau ;\varepsilon)}{\partial \tau}+\O(\varepsilon^3)\,,
\end{align}
\text{and}
\begin{align}
    &\varepsilon \frac{\varphi(t,\tau;\varepsilon) }{6} \int_{t_i}^{t} \d t' \,
	    v_{\rm R}(t-t')\varphi ^2(t',\tau ';\varepsilon) \notag \\
    &=\varepsilon \frac{\varphi(t,\tau;\varepsilon) }{6} \int_{t_i}^{t} \d t' \, v_{\rm R}(t-t')\varphi ^2(t',\tau ;\varepsilon) \notag\\
    &\qquad\qquad\qquad\qquad- \varepsilon^2 \frac{\varphi(t,\tau;\varepsilon) }{3} \int_{t_i}^{t} \d t' \, (t-t') v_{\rm R}(t-t')\varphi (t',\tau ;\varepsilon)\frac{\partial \varphi(t',\tau ;\varepsilon)}{\partial \tau}+\O(\varepsilon^3)\,.
\end{align}
\label{eq:non-local terms in powers of varepsilon}
\end{subequations}

Fourth, with Eq.~\eqref{eq:varphi in powers of varepsilon} and Eq.~\eqref{eq:non-local terms in powers of varepsilon}, one can organize Eq.~\eqref{eq:condensate eom4} in powers of $\varepsilon$. The leading order is given by
\begin{equation}\label{eq:eom varphi_0}
    \frac{\partial ^2 \varphi_0(t,\tau) }{\partial t^2} + M_\phi^2 \varphi_0(t,\tau)=0\,,
\end{equation}
which is a harmonic oscillator equation. 
The subleading equation is 
\begin{align}
   \frac{\partial ^2 \varphi_1(t,\tau) }{\partial t^2} + M_\phi^2 \varphi_1(t,\tau) =&-2  \frac{\partial ^2 \varphi_0(t,\tau) }{\partial t \partial \tau}-3 H(\tau) \frac{\partial \varphi_0(t,\tau) }{\partial t}  - \frac{\lambda_\phi }{6} \varphi_0 ^3(t,\tau) \notag \\
&-\int_{t_i}^{t} \d t' \, \pi_{\rm R}(t-t') \varphi_0(t',\tau)-\frac{\varphi_0(t,\tau) }{6} \int_{t_i}^{t} \d t' \, v_{\rm R}(t-t')\varphi_0 ^2(t',\tau)\,.
\label{eq:eom varphi_1_0}
\end{align}
The solution for the harmonic oscillator equation \eqref{eq:eom varphi_0} has the form
\begin{equation}
\label{eq:formvarphi0}
    \varphi_0(t,\tau)=\text{Re}\left[R(\tau) \e^{-\i M_\phi t} \right]\,.
\end{equation}
Note that the above solution is obtained when $M_\phi$ is constant. However, $M_\phi$ contains thermal corrections which are temperature dependent and are necessarily time dependent if one considers an expanding universe. Therefore, the multi-scale analysis used here is valid only when $M_\phi$ is dominated by $m^2_\phi$, i.e., $M_\phi\approx m_\phi$. In this case, one can actually take  $(M^2_\phi-m^2_\phi)\varphi$ as an additional perturbation term in the equation of motion. Here we still treat $M^2_\phi \varphi$ altogether as a term so that the solutions for an expanding universe can be compared directly with those for a flat spacetime given in Ref.~\cite{Ai:2021gtg}. One shall bear in mind that the solutions for expanding universes are valid only for $M_\phi\approx m_\phi$ while for flat spacetime there is no such constraint.

Note that the solution $\varphi_0$ has not been determined yet since $R(\tau)$ is unknown. To obtain $R(\tau)$, one has to look into the equation of motion at the next-leading order.
With the form~\eqref{eq:formvarphi0},  the non-local terms in Eq.~\eqref{eq:eom varphi_1_0} can be written as
\begin{align}
\label{eq:non-local-fourier_11}
  -\int_{t_i}^{t} \d t' \, \pi_{\rm R}(t-t') \varphi_0(t',\tau) =&-\text{Re}\left[R(\tau) \e^{-\i M_\phi t} \int_{t_i}^{t} \d t' \, \pi_{\rm R}(t-t') \e^{\i M_\phi (t-t')}\right]\,,
\end{align}
and
\begin{align}
\label{eq:non-local-fourier_12}
-\frac{\varphi_0(t,\tau) }{6} \int_{t_i}^{t} \d t' \, v_{\rm R}(t-t')\varphi_0 ^2(t',\tau) =&-\frac{1}{12}\text{Re}\left[\e^{-\i M_\phi t}[R(\tau)]^2 R^*(\tau)  \int_{t_i}^{t} \d t' \, v_{\rm R}(t-t') \right]\notag \\
&-\frac{1}{24}\text{Re}\left[\e^{-\i M_\phi t} [R(\tau)]^2 R^*(\tau)  \int_{t_i}^{t} \d t' \, v_{\rm R}(t-t') \e^{2\i M_\phi (t-t')}\right]\notag \\
&-\frac{1}{24}\text{Re}\left[\e^{-3\i M_\phi t} [R(\tau)]^3  \int_{t_i}^{t} \d t' \, v_{\rm R}(t-t') \e^{2\i M_\phi (t-t')}\right]\,.
\end{align}
The integrals on the RHS of Eqs.~\eqref{eq:non-local-fourier_11} and~\eqref{eq:non-local-fourier_12} have the same form as Fourier transforms except for the upper and lower limits. Following Ref.~\cite{Ai:2021gtg}, we make
the approximation of neglecting any early-time transient effects due to initial conditions in the non-local terms. This allows us to take $t_i \rightarrow -\infty$, eliminating the explicit finite $t_i$ dependence appearing in the lower limit of the integrals.
On the other hand, the appearance of the Heaviside step function in the definitions of the retarded self-energy and the retarded four-vertex function (cf., Eq.~\eqref{piR+vR def}) allows us to replace $t$ by $+\infty$ in the upper limit of the integrals. 

So, we could actually recognize the integrals on the RHS of Eqs.~\eqref{eq:non-local-fourier_11} and~\eqref{eq:non-local-fourier_12} as Fourier transforms. More explicitly, we have
\begin{subequations}
\label{eq:FourierTransf}
\begin{align}
    \int_{t_i}^{t} \d t' \, \pi_{\rm R}(t-t') \e^{\i M_\phi (t-t')} &\approx \int_{-\infty}^{+\infty} \d t' \, \pi_{\rm R}(t-t') \e^{\i M_\phi (t-t')}=\widetilde{\pi}_{\rm R} (M_\phi )\,,\\
    \int_{t_i}^{t} \d t' \, v_{\rm R}(t-t') &\approx \int_{-\infty}^{+\infty} \d t' \, v_{\rm R}(t-t') \e^{\i 0(t-t')}=\widetilde{v}_{\rm R} (0)\,,\\
\int_{t_i}^{t} \d t' \, v_{\rm R}(t-t') \e^{2\i M_\phi (t-t')} &\approx \int_{-\infty}^{+\infty} \d t' \, v_{\rm R}(t-t') \e^{2\i M_\phi (t-t')}=\widetilde{v}_{\rm R} (2M_\phi )\,.
\end{align}
\end{subequations}
Therefore, the subleading equation~\eqref{eq:eom varphi_1_0} now reads
\begin{align}
     \frac{\partial ^2 \varphi_1(t,\tau) }{\partial t^2} +& M_\phi^2 \varphi_1(t,\tau) =2\text{Re}\bigg\{ \i M_\phi \e^{-\i M_\phi t} \Big[ \frac{\d R(\tau)}{\d \tau}+\frac{3}{2}H(\tau) R(\tau)+\frac{\i \lambda_\phi}{16M_\phi}[R(\tau)]^2 R^*(\tau) \notag\\
&+\frac{\i R(\tau)}{2M_\phi}\widetilde{\pi}_{\rm R} (M_\phi )+\frac{\i }{24 M_\phi}\widetilde{v}_{\rm R} (0 )[R(\tau)]^2 R^*(\tau)+\frac{\i }{48 M_\phi}\widetilde{v}_{\rm R} (2M_\phi )[R(\tau)]^2  R^*(\tau)\Big]\bigg\}\notag\\
&+2\text{Re}\bigg\{ \i M_\phi \e^{-\i 3M_\phi t}\Big[\frac{\i \lambda_{\phi}}{48 M_\phi} +\frac{\i}{48M_{\phi}} \widetilde{v}_{\rm R} (2M_\phi )\Big][R(\tau)]^3\bigg\}\,.
\label{eq:eom varphi_1_1}
\end{align}
Defining
\begin{align}
\label{eq:important_quantities}
\mu&\equiv \frac{\text{Re}[\widetilde{\pi}_{\rm R} (M_\phi )]}{2M_{\phi}}\,, \; \gamma \equiv -\frac{\text{Im} [\widetilde{\pi}_{\rm R} (M_\phi )]}{2M_{\phi}}\,, \; \sigma \equiv -\frac{\text{Im} [\widetilde{v}_{\rm R} (2M_\phi )]}{24 M_{\phi}}
\,,\notag\\
\alpha_1 &\equiv \frac{\text{Re} [\widetilde{v}_{\rm R} (0)]}{24 M_{\phi}}\,,\ \ \, \alpha _2\equiv \frac{\text{Re} [\widetilde{v}_{\rm R} (2M_\phi )]}{48 M_{\phi}}\,,\ \; \alpha \equiv \alpha_1+\alpha_2\,,
\end{align}
Eq.~\eqref{eq:eom varphi_1_1} can be written as 
\begin{align}
     \frac{\partial ^2 \varphi_1(t,\tau) }{\partial t^2} + M_\phi^2 \varphi_1(t,\tau) 
=&2\text{Re}\Bigg\{ \i M_\phi \e^{-\i M_\phi t} \bigg[ \frac{\d R(\tau)}{\d \tau}+\Big(\frac{3}{2}H(\tau)+ \i \mu +\gamma \Big)R(\tau)\notag\\
&+\Big(\frac{\i \lambda_\phi}{16M_\phi}+\i \alpha +\frac{\sigma}{2}\Big) [R(\tau)]^2 R^*(\tau) \bigg]\Bigg\}\notag\\
&+2\text{Re}\Bigg\{ \i M_\phi \e^{-\i 3M_\phi t}\bigg(\frac{\i \lambda_{\phi}}{48 M_\phi}+ \i \alpha_2 +\frac{\sigma}{2}\bigg)[R(\tau)]^3\Bigg\}\,,
\label{eq:eom varphi_1_2}
\end{align}
where we used  
\begin{equation}
    \text{Im}[\widetilde{v}_{\rm R}(0)]=0\,,
\end{equation}
which is due to that the imaginary part of the Fourier-transformed proper four-vertex is anti-symmetric. 

Finally, we determine $R(\tau)$ from Eq.~\eqref{eq:eom varphi_1_2}.  We have organized the equation of motion for $\varphi_1 (t,\tau)$ in such a way that it describes the evolution of a harmonic oscillator $\varphi_1 (t,\tau)$  driven by two oscillating forces.  One force described by the first two lines on the RHS of Eq.~\eqref{eq:eom varphi_1_2} has a frequency $M_\varphi$, which is the same as the natural frequency of  $\varphi_1 (t,\tau)$. The other force given by the last line has a frequency $3 M_\varphi$. The second force just modifies the original oscillating behavior by adding different oscillations with constant amplitudes.  However, it is well known that the first force can lead to a resonant behavior where the final amplitude of $\varphi_1 (t,\tau)$ would be proportional to the time $t$, i.e., the amplitude of $\varphi_1 (t,\tau)$ will increase without bound. As a standard procedure of the multi-scale analysis~\cite{Bender}, to avoid the non-physical spurious resonances we require that 
\begin{equation}
\label{eq:eom R_1}
    \frac{\d R(\tau)}{\d \tau}+\Big(\frac{3}{2}H(\tau)+ \i \mu +\gamma \Big)R(\tau) +\Big(\frac{\i \lambda_\phi}{16M_\phi}+\i \alpha +\frac{\sigma}{2}\Big) [R(\tau)]^2 R^*(\tau)=0\,
\end{equation}
such that the first force vanishes. Making the following Ansatz 
\begin{equation}
    R(\tau) = A (\tau) \,\e ^{-\i f(\tau)}\,,
\end{equation}
where $A (\tau)$ and $f(\tau)$ are real functions of $\tau$, we can split the complex equation~\eqref{eq:eom R_1} into the following two real equations,
\begin{subequations}
\label{eq:eom_A and f(tau)}
\begin{align}
\label{eq:eom_A(tau)}
    \frac{\d A (\tau)}{\d \tau}+\left(\gamma +\frac{3}{2}H(\tau)\right) A(\tau) +\frac{\sigma}{2} [A(\tau)]^3&=0\,,\\
    \frac{\d f(\tau)}{\d \tau}-\mu -\left(\frac{\lambda_{\phi}}{16M_{\phi}}+\alpha \right)[A(\tau)]^2&=0\,.
\end{align}
\end{subequations}
The solution for $A(\tau)$ depends on $\gamma$, $H(\tau)$ and $\sigma$. Other than these three parameters,  the solution for $f(\tau)$ also depends on $\mu$, $\lambda_{\phi}$, and $\alpha$.

To be general, we parameterize the Hubble constant as
\begin{equation}
    H(\tau) = \frac{\zeta}{\tau}\,,
\end{equation}
where $\zeta$ is a non-negative real number. Typical universes with different $\zeta$ will be discussed below. 
The solution of $A(\tau)$ can be written as
\begin{align}
\label{eq:sol A(t)_0}
    A(\tau) = \frac{\sqrt{2  }}{\e^{ \gamma \tau }(2 \tau )^{\frac{3\zeta}{2}}\sqrt{- \gamma  ^{3 \zeta-1}\sigma   \, \Gamma (1-3 \zeta,2  \gamma \tau )+ c_0  }} \,,
\end{align}
where $c_0$ is a constant of integration and $\Gamma (a,x)$ is the incomplete gamma function defined by
\begin{equation}
    \Gamma (a,x)=\int^{+\infty}_{x}\d y\, \e^{-y} y^{a-1}\,.
\end{equation}
The leading approximation for the solution of Eq.~\eqref{eq:condensate eom2} then reads
\begin{align}
\label{eq:sol_varphi_0}
    \varphi (t)\approx \varphi_0 (t,t) 
    =\frac{\sqrt{2  }}{\e^{ \gamma t }(2 t )^{\frac{3\zeta}{2}}\sqrt{- \gamma  ^{3 \zeta-1}\sigma   \, \Gamma (1-3 \zeta,2  \gamma t )+ c_0 }} \times \cos\left[f(t)+M_{\phi}t\right]\,.
\end{align}
The solution of $f(t)$ cannot be expressed by fundamental functions for unspecified $\zeta$. It is already clear from Eq.~\eqref{eq:sol_varphi_0} that the expansion of the universe ($\zeta>0$) induces a power-law damping behavior for the condensate. The leading approximation of the energy density for the condensate $\varphi$ is given by
\begin{equation}
   \rho _{\varphi}= \frac{1}{2} \dot{\varphi}_0 ^2 +\frac{1}{2} M^2_{\phi} \varphi _0 ^2 \,.
\end{equation}
This quantity will be used for the discussion of particle production. In what follows, we shall study the solution Eq.~\eqref{eq:sol_varphi_0} in different situations.

\subsection{Static universe}

Taking $\zeta=0$ gives a static universe. For future use, we discuss separately the cases when $\gamma$ (the imaginary part of the self-energy) is negligible and when not. The reason for doing this is that $\gamma$ and $\sigma$ have very different effects on the evolution of the condensate in an expanding universe. In flat spacetime or a static universe, regardless of whether $\gamma$ is vanishing or not, the decay of the condensate is always complete. We will see that the conclusion is completely different in an expanding universe. As $\gamma$ is suppressed in the low temperature (see Eqs.~\eqref{eq:Im-pi} and~\eqref{eq:Li2-lowT}) while $\sigma$ is nonvanishing even at zero temperature, there could be some range for the temperature in which $\gamma$ is negligible. In such a case, we simply take $\gamma=0$.

\subsubsection{Non-negligible \texorpdfstring{$\gamma$}{TEXT}}

We first consider the case of non-negligible $\gamma$. In this case, we have the solution of $A(t)$,
\begin{align}
\label{eq:sol varphi_static universe1}
A(t)
= \frac{A_0 \e^{-\gamma t}}{\sqrt{1+\frac{\sigma A_0^2}{2\gamma}(1-\e^{-2\gamma t})}}\,, 
\end{align}
where we have used the identity $\Gamma (1,x) = \e^{-x}$ and $A_0$ is related to $c_0$ via $1/A_0^2+\sigma/(2\gamma)=c_0/2$. The solution of $f(t)$ reads
\begin{align}
\label{eq:sol varphi_static universe2}
f(t)=f_0+\mu t +\frac{ (\lambda_{\phi} +16 \alpha  M_{\phi}) \ln \left[1+\frac{\sigma A^2 _0}{2 \gamma}(1-\e^{-2\gamma t})\right]}{16 M_{\phi} \sigma}\,.
\end{align}
The constants of integration are determined by the initial conditions for the condensate. Note that our solution given by Eqs.~\eqref{eq:sol varphi_static universe1} and~\eqref{eq:sol varphi_static universe2} agrees with the result in Ref.~\cite{Ai:2021gtg}.\footnote{See Eq.(3.92) of that reference. Note that the $\gamma$ defined in our paper is one half of that in Ref.~\cite{Ai:2021gtg}.}

The energy density of the oscillating scalar field can be approximated as 
\begin{align}
\label{eq:rho static universe_0}
    \rho _{\varphi} 
    =&\frac{1}{2} \left[ \dot{A}(t)\cos(f+M_{\phi}t)-A(t)\sin(f+M_{\phi}t)(M_{\phi}+\dot{f})\right]^2+\frac{1}{2}M_{\phi}^2 [A(t)]^2 [\cos(f+M_{\phi}t)]^2 \notag\\
    \approx & \frac{1}{2} M_{\phi}^2 [A(t)]^2
    \approx  \frac{A^2 _0 M^2 _{\phi}}{2}\cdot \frac{\e ^{-2\gamma t}}{1+\frac{\sigma A^2 _0}{2\gamma}(1-\e ^{-2\gamma t})} \,.
\end{align}
To get the second line in Eq.~\eqref{eq:rho static universe_0}, we have used the assumption that
\begin{align}\label{eq:static universe assumption_0}
   \frac{\dot{A}}{A} \ll M_{\phi} \,,\;\; \frac{\d f}{\d t}\ll M_{\phi}\,,
\end{align}
which implies that the amplitude and the frequency of oscillations do not change much during one oscillation. From Eq.~\eqref{eq:rho static universe_0}, we see that the energy density of the condensate experiences first  power-law damping [$1/(1+\sigma A^2 _0 t)$] at early times and then dominantly exponential damping at late times. More explicitly, we have 
\begin{align}
\label{eq:drho_dt_static}
    \frac{1}{\rho_\varphi}\frac{\d \rho_{\varphi}}{ \d t} \approx -2\gamma -\sigma [A(t)]^2 \,,
\end{align}
from which we observe that the decay rate of the energy density $\rho_{\varphi}$ can be separated into two parts; one is a constant proportional to $\gamma$ and the other one is proportional to $\sigma$ which falls off as $[A(t)]^2$. At late times, the decay rate is dominated by the constant term. 

\subsubsection{Microscopic interpretation}
\label{sec:micro-inter}

The damping of the condensate oscillations and the decrease of its energy density can be interpreted as particle production. Actually, the self-energy corresponds to the following diagrams in the effective action,
\begin{align}
\label{eq:diag-self-energy}
    \begin{tikzpicture}[baseline={0cm-0.5*height("$=$")}]
\draw[thick] (0,0) circle (0.5) ;
\draw[thick] (-0.5,0) -- (0.5,0) ;
\filldraw (-0.5,0) circle (1.5pt) node {} ;
\draw[thick] (0.5,0) -- (1,0) ;
\draw[thick] (1.15,0) circle (0.15) ;
\draw[thick] (1.15-0.707*0.15,0.707*0.15) -- (1.15+0.707*0.15,-0.707*0.15) ;
\draw[thick] (1.15-0.707*0.15,-0.707*0.15) -- (1.15+0.707*0.15,+0.707*0.15) ;
\filldraw (0.5,0) circle (1.5pt) node {} ;
\draw[thick] (-1,0) -- (-0.5,0) ;
\draw[thick] (-1.15,0) circle (0.15) ;
\draw[thick] (-1.15-0.707*0.15,0.707*0.15) -- (-1.15+0.707*0.15,-0.707*0.15) ;
\draw[thick] (-1.15-0.707*0.15,-0.707*0.15) -- (-1.15+0.707*0.15,+0.707*0.15) ;
\end{tikzpicture}\,,
\quad
\begin{tikzpicture}[baseline={0cm-0.5*height("$=$")}]
\draw[thick,dashed] (0,0) circle (0.5) ;
\draw[thick] (-0.5,0) -- (0.5,0) ;
\draw[thick] (0.5,0) -- (1,0) ;
\draw[thick] (1.15,0) circle (0.15) ;
\draw[thick] (1.15-0.707*0.15,0.707*0.15) -- (1.15+0.707*0.15,-0.707*0.15) ;
\draw[thick] (1.15-0.707*0.15,-0.707*0.15) -- (1.15+0.707*0.15,+0.707*0.15) ;
\filldraw[fill=black] (0.5,0) circle (1.5pt) node {} ;
\draw[thick] (-1,0) -- (-0.5,0) ;
\draw[thick] (-1.15,0) circle (0.15) ;
\draw[thick] (-1.15-0.707*0.15,0.707*0.15) -- (-1.15+0.707*0.15,-0.707*0.15) ;
\draw[thick] (-1.15-0.707*0.15,-0.707*0.15) -- (-1.15+0.707*0.15,+0.707*0.15) ;
\filldraw[fill=black] (-0.5,0) circle (1.5pt) node {} ;
\end{tikzpicture}\,,
\end{align}
where a line ended with a wheel cross denotes the scalar background $\varphi$.
If $\varphi$ is a perturbative particle, the cutting rules~\cite{Cutkosky:1960sp,Weldon:1983jn,Kobes:1985kc,Kobes:1986za,Landshoff:1996ta,Gelis:1997zv,Bedaque:1996af} suggest that processes contributing to a non-vanishing imaginary part of the retarded self-energy consist of $\varphi\leftrightarrow \phi\phi\phi$, $\varphi\leftrightarrow \phi\chi\chi$ and rearrangements thereof (e.g., $\varphi\chi \leftrightarrow \phi\chi$). However, the condensate quanta have exactly zero momentum and their energy are equal to the mass of $\phi$-particles, i.e., every $\varphi$ quantum has four-momentum $(M_\phi,\vec{0})$. Thus some processes cannot satisfy the on-shell conditions. Apparently, $\varphi\leftrightarrow \phi\phi\phi$, $\varphi\leftrightarrow \phi\chi\chi$ are not kinematically possible. But the processes $\varphi\chi\chi\leftrightarrow \phi$, $\varphi\phi\chi\leftrightarrow \chi$, $\varphi\phi\phi\leftrightarrow\phi$ also cannot be on-shell.\footnote{A detailed derivation for the self-energy (e.g., Eq.~\eqref{eq:Im-pi}) that makes this statement more apparent will be given in Ref.~\cite{AL}.} The processes contributing to the $\gamma$ dissipation are therefore 
\begin{align}
    \gamma{\rm \ channels:}\quad \label{eq:decaychannel-one}
    \varphi\phi\leftrightarrow \phi\phi\,,\quad
    \varphi\chi\leftrightarrow \phi\chi\,,\quad \varphi\phi\leftrightarrow\chi\chi\,.
\end{align}
These are the condensate decay channels with one condensate quantum. We call them the {\it $\gamma$ channels}. Note that the dissipation of the condensate is a total effect of the decay processes and their inverses. These ``scattering'' processes are possible at finite temperature because the plasma contains many $\chi$- and $\phi$-particles. The damping caused by them are called {\it Landau damping}~\cite{Bellac:2011kqa}. The $\gamma$ channels are absent at zero temperature.

The diagrams in the effective action corresponding to the proper four-vertex function are
\begin{align}
    \begin{tikzpicture}[baseline={0cm-0.5*height("$=$")}]
\draw[thick] (0,0) circle (0.5) ;
\filldraw (0.5,0) circle (1.5pt) node {} ;
\filldraw (-0.5,0) circle (1.5pt) node {} ;
\draw[thick] (0.5,0) -- (0.6+0.3*0.707,0.1+0.3*0.707) ;
\draw[thick] (0.6+0.45*0.707,0.1+0.45*0.707) circle (0.15) ;
\draw[thick] (0.6+0.3*0.707,0.1+0.3*0.707) -- (0.6+0.6*0.707,0.1+0.6*0.707) ;
\draw[thick] (0.6+0.6*0.707,0.1+0.3*0.707) -- (0.6+0.3*0.707,0.1+0.6*0.707) ;
\draw[thick] (0.5,0) -- (0.6+0.3*0.707,-0.1-0.3*0.707) ;
\draw[thick] (0.6+0.45*0.707,-0.1-0.45*0.707) circle (0.15) ;
\draw[thick] (0.6+0.3*0.707,-0.1-0.3*0.707) -- (0.6+0.6*0.707,-0.1-0.6*0.707) ;
\draw[thick] (0.6+0.6*0.707,-0.1-0.3*0.707) -- (0.6+0.3*0.707,-0.1-0.6*0.707) ;
\draw[thick] (-0.5,0) -- (-0.6-0.3*0.707,-0.1-0.3*0.707) ;
\draw[thick] (-0.6-0.45*0.707,-0.1-0.45*0.707) circle (0.15) ;
\draw[thick] (-0.6-0.3*0.707,-0.1-0.3*0.707) -- (-0.6-0.6*0.707,-0.1-0.6*0.707) ;
\draw[thick] (-0.6-0.6*0.707,-0.1-0.3*0.707) -- (-0.6-0.3*0.707,-0.1-0.6*0.707) ;
\draw[thick] (-0.5,0) -- (-0.6-0.3*0.707,0.1+0.3*0.707) ;
\draw[thick] (-0.6-0.45*0.707,0.1+0.45*0.707) circle (0.15) ;
\draw[thick] (-0.6-0.3*0.707,0.1+0.3*0.707) -- (-0.6-0.6*0.707,0.1+0.6*0.707) ;
\draw[thick] (-0.6-0.6*0.707,0.1+0.3*0.707) -- (-0.6-0.3*0.707,0.1+0.6*0.707) ;
\end{tikzpicture}\,,
\quad
\begin{tikzpicture}[baseline={0cm-0.5*height("$=$")}]
\draw[thick, dashed] (0,0) circle (0.5) ;
\filldraw (0.5,0) circle (1.5pt) node {} ;
\filldraw (-0.5,0) circle (1.5pt) node {} ;
\draw[thick] (0.5,0) -- (0.6+0.3*0.707,0.1+0.3*0.707) ;
\draw[thick] (0.6+0.45*0.707,0.1+0.45*0.707) circle (0.15) ;
\draw[thick] (0.6+0.3*0.707,0.1+0.3*0.707) -- (0.6+0.6*0.707,0.1+0.6*0.707) ;
\draw[thick] (0.6+0.6*0.707,0.1+0.3*0.707) -- (0.6+0.3*0.707,0.1+0.6*0.707) ;
\draw[thick] (0.5,0) -- (0.6+0.3*0.707,-0.1-0.3*0.707) ;
\draw[thick] (0.6+0.45*0.707,-0.1-0.45*0.707) circle (0.15) ;
\draw[thick] (0.6+0.3*0.707,-0.1-0.3*0.707) -- (0.6+0.6*0.707,-0.1-0.6*0.707) ;
\draw[thick] (0.6+0.6*0.707,-0.1-0.3*0.707) -- (0.6+0.3*0.707,-0.1-0.6*0.707) ;
\draw[thick] (-0.5,0) -- (-0.6-0.3*0.707,-0.1-0.3*0.707) ;
\draw[thick] (-0.6-0.45*0.707,-0.1-0.45*0.707) circle (0.15) ;
\draw[thick] (-0.6-0.3*0.707,-0.1-0.3*0.707) -- (-0.6-0.6*0.707,-0.1-0.6*0.707) ;
\draw[thick] (-0.6-0.6*0.707,-0.1-0.3*0.707) -- (-0.6-0.3*0.707,-0.1-0.6*0.707) ;
\draw[thick] (-0.5,0) -- (-0.6-0.3*0.707,0.1+0.3*0.707) ;
\draw[thick] (-0.6-0.45*0.707,0.1+0.45*0.707) circle (0.15) ;
\draw[thick] (-0.6-0.3*0.707,0.1+0.3*0.707) -- (-0.6-0.6*0.707,0.1+0.6*0.707) ;
\draw[thick] (-0.6-0.6*0.707,0.1+0.3*0.707) -- (-0.6-0.3*0.707,0.1+0.6*0.707) ;
\end{tikzpicture}\,.
\end{align}
Similarly, the cutting rules suggest that processes contributing to the $\sigma$ dissipation coefficient (due to a non-vanishing imaginary part of the retarded proper four-vertex function) consist of
\begin{align}
\label{eq:decaychannel-two}
\sigma\ {\rm channel:}\quad
\quad (\varphi\varphi)\leftrightarrow \chi\chi\quad  {\rm for\ } M_\phi> M_\chi\,.
\end{align}
Note that the processes $(\varphi\varphi)\phi\leftrightarrow \phi$, $(\varphi\varphi)\chi\leftrightarrow\chi$ cannot satisfy the on-shell conditions because the $\phi$ particles (or $\chi$ particles) on the LHS and RHS would have the same momentum but different energy and thus cannot be on-shell simutaneously. The process $(\varphi\varphi)\leftrightarrow \phi\phi$ is on the edge of satisfying the on-shell conditions. The contribution from this process is given by the first term of Eq.~\eqref{Im vR omega T} when taking $\omega=2M_\phi$, and thus vanishes also. The process in Eq.~\eqref{eq:decaychannel-two} is the decay channel with two condensate quanta and we call it the {\it $\sigma$ channel}. It survives at zero temperature.

\subsubsection{Negligible \texorpdfstring{$\gamma$}{TEXT}}
\label{sec:gamma=0}

Now we consider the case where the imaginary part of the self-energy is negligible. This would mean that we take $\gamma =0$ in Eq.~\eqref{eq:eom_A and f(tau)} ($\mu$ in general is not vanishing). Then we obtain the following solutions for $A(t)$ and $f(t)$,
   \begin{subequations}\label{eq:sol varphi_static universe1_gamma0}
   \begin{align}
       A(t)&=\frac{1}{\sqrt{\sigma t+c_0}}\,,\\
       f(t)&=f_0+\mu t+\frac{(\lambda_{\phi} +16 \alpha  M_{\phi}) \ln ( c_0 + \sigma  t)}{16 M_{\phi} \sigma }\,,
   \end{align}
    \end{subequations}
where $c_0$ and $f_0$ are constants of integration. Here and in what follows, $c_0$ and $f_0$ in the solutions for different situations are not necessarily identical. An example of this solution is shown by the brown line in Fig.~\ref{fig:condensate-radiation universe_gamma0_ti=100(new)}.  

\begin{figure}[ht]
\centering
\includegraphics[scale=1]{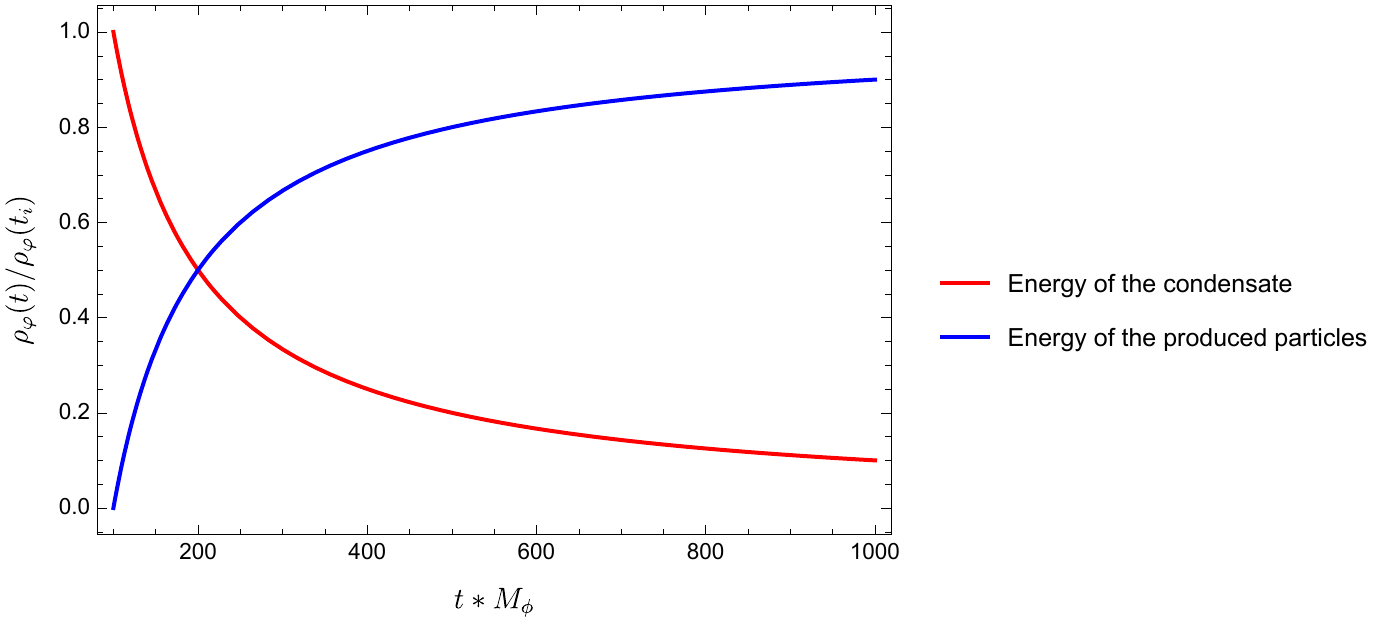}
\caption{Energy transfer from the condensate to the produced particles in a static universe without the self-energy correction. The evolution of the condensate energy density and the produced particles are shown by the red and blue lines, respectively. We take $\sigma =0.01 / M_{\phi}$ and the initial conditions at $t_i M_{\phi}= 100$ as $\varphi(t_i) /M_{\phi}=1 $ and $\dot{\varphi} (t_i)=0$ (the same initial conditions are used in all the plots below). The energy of the condensate can be fully transferred to the produced particles.} 
\label{comovingenergy_h0_gamma0_ti=100}
\end{figure}

The energy density of the oscillating scalar field can be approximated as 
\begin{align}\label{eq:rho static universe_1}
    \rho _{\varphi} 
    \approx  \frac{1}{2} M_{\phi}^2 [A(t)]^2
    \approx  \frac{ M^2 _{\phi}}{2(\sigma t+ c_0)} \,.
\end{align}
where again we have assumed that the amplitude and the frequency of oscillations do not change much during one oscillation. The energy density of the condensate goes to $0$ when $t \to \infty$, see Fig.~\ref{comovingenergy_h0_gamma0_ti=100}. The energy transfer from the condensate to the produced particles is efficient even if there are only the $\sigma$ channel.  However, as we shall see shortly, the picture will be totally different if an expanding universe is considered.

\subsection{Radiation-dominated universe}
\label{sec:Radiation-dominated universe}

Let us now discuss a radiation-dominated universe for which $\zeta=1/2$. This is the most relevant situation as we are interested in the radiation-dominated period after preheating. To have a comprehensive understanding of particle production in an expanding universe, below we discuss three different cases. 

\subsubsection{No interactions}

The evolution of the condensate without interactions in a radiation-dominated universe can be obtained by setting $\gamma =\sigma=\lambda_{\phi}=\mu =\alpha=0$ and $\zeta=1/2$.  Then, the leading approximation for the solution of $\varphi (t)$ is given by
\begin{align}
\label{eq:varphi radiation no interation}
\varphi(t)\approx \frac{1}{ 2^{1/4} \sqrt{  c_0  }}\,t^{-3/4}\times \cos\left(f_0+M_{\phi}t\right)\,,
\end{align}
where $c_0$ and  $f_0$ are constants of integration. In this situation, the oscillation amplitude decreases as $[a(t)]^{-3/2}$ with $a(t) \propto \sqrt{t}$. This damping is simply due to the expansion of the universe. The total energy of the condensate in a given physical volume $V$ is given by
\begin{equation}
    \rho _{\varphi} V \approx  \frac{V}{2}M_{\phi} ^2 A^2\,.
\end{equation}
Note that in a spatially flat FLRW universe, the physical volume is given by $V = a^3\, V_{\text{c}}$ where $V_{\text{c}}$ is the constant comoving volume. Hence, the total energy of the condensate is reflected by the quantity $a^3 \rho _{\varphi}\equiv \tilde\rho_\varphi$ to which we refer as the {\it comoving energy density} in this paper. Since $A (t) \propto [a(t)]^{-3/2}$, we have $\rho _{\varphi} \propto a^{-3}$ and
\begin{align}
    \tilde{\rho}_\varphi \equiv a^3 \rho _{\varphi} = \text{constant}\,,
\end{align}
which implies that the total energy of the condensate is conserved. 

Below, we shall discuss situations with interactions in which particle production would occur and the condensate comoving energy density is not conserved anymore. We assume that the produced particles are relativistic (this could be the case if the plasma temperature is high compared to all the thermal masses). Let $\rho_{\rm pp}$ denote the energy density of the produced particles. Then the total energy for the produced particles is characterized by  $\tilde{\rho}_{\rm pp}\equiv a^4\rho_{\rm pp}$. Without interactions, $\tilde{\rho}_{\rm pp}$ would be a conserved quantity. With interactions, we have the following relation between $\tilde{\rho}_{\rm pp}$ and $\tilde{\rho}_\varphi$,
\begin{align}
   \dot{\rho}_{\rm pp}+4H\rho_{\rm pp}=-\left(\dot{ \rho}_\varphi +3 H\rho_{\varphi}\right)\quad \Rightarrow\quad \quad \frac{\d \tilde{\rho}_{\rm pp}}{\d t}=-a\,\frac{\d \tilde{\rho}_\varphi}{\d t}\,.
\end{align}
Thus knowing the evolution of the condensate comoving energy density $\tilde{\rho}_\varphi$, one can integrate the above equation to obtain $\tilde{\rho}_{\rm pp}$. To avoid assuming some constants in $a(t)$ for different universes, below when discussing the energy transfer from the condensate to the produced particles, we simply compare the condensate comoving energy density with the energy loss defined as
\begin{align}
    \tilde{\rho}_{\rm loss}(t)\equiv \tilde{\rho}_\varphi(t_i)-\tilde{\rho}_\varphi(t)\,.
\end{align}

Note that the amplitude $A(t)$ in Eq.~\eqref{eq:varphi radiation no interation} is divergent at $t\to 0^+$, which comes from the divergent behaviour of the Hubble parameter at $t\to 0^+$. This is not surprising since the time $t$ in Eq.~\eqref{eq:varphi radiation no interation} is the cosmic time and $t \to 0^+$ corresponds to the so-called big-bang singularity. Thus, the solution for the condensate in an expanding universe should be applied for times larger than zero. 

\subsubsection{Negligible \texorpdfstring{$\gamma$}{TEXT}}

\label{sec:radiation_gamma0}

If we take $\gamma=0$, the solutions of $A(t)$ and $f(t)$ are given by\footnote{The relation $\Gamma(-1/2,x)\rightarrow 2/\sqrt{x}$ as $x\rightarrow 0$ is needed if one derives the solution for $A(t)$ by taking the limit $\gamma\rightarrow 0$ in Eq.~\eqref{eq:sol_varphi_0}.}
\begin{subequations}
\label{eq:sol-varphi-gamma=0_c=1/2}
\begin{align}
A(t)&=\frac{1}{\sqrt{c_0\,t^{3/2}-2\sigma t}}\,,\\
f(t)&=f_0+\mu t-\frac{(\lambda_{\phi} +16 \alpha  M_{\phi}) }{32 M_{\phi} \sigma } \left[\ln (t)-2 \ln \left(c_0 \sqrt{t}-2 \sigma \right)\right]\,,
\end{align}
\end{subequations}
where $c_0$ and $f_0$ are constants of integration.

\begin{figure}[ht]
\centering
\includegraphics[scale=1]{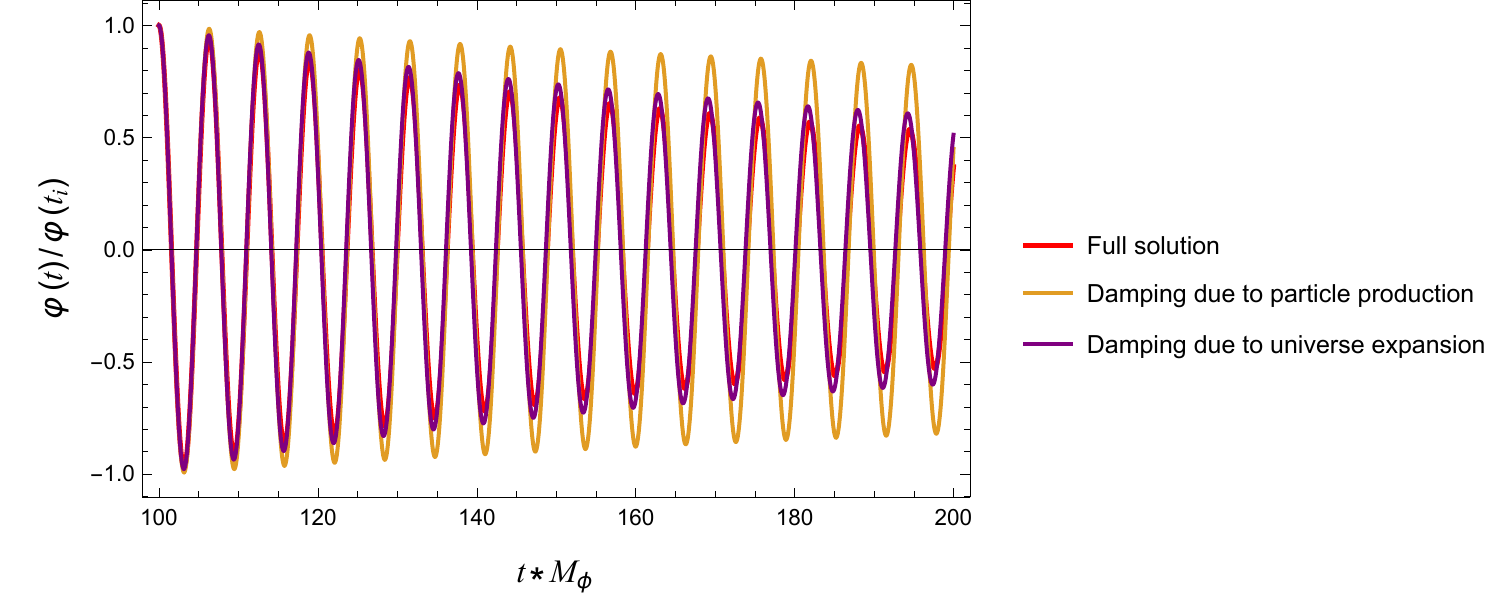}
\caption{Solutions of the condensate equation \eqref{eq:condensate eom2} with the imaginary part of the self-energy neglected ($\gamma=0$). Analytic approximation in a radiation-dominated universe, i.e., Eqs.~\eqref{eq:sol-varphi-gamma=0_c=1/2}, is drawn in the red line with $\alpha=-0.01 /M_{\phi}$, $\sigma=0.005/M_{\phi}$, $\mu =0.001 M_{\phi}$, and $\lambda_{\phi}=0.05$. Analytic approximation in a static universe with the same parameters, given by Eqs.~\eqref{eq:sol varphi_static universe1_gamma0}, is drawn in the brown line. Analytic approximation without interactions in a radiation-dominated universe, i.e., Eq.~\eqref{eq:varphi radiation no interation}, is drawn in the purple line. }
\label{fig:condensate-radiation universe_gamma0_ti=100(new)}
\end{figure}

\begin{figure}[ht]
\centering
\includegraphics[scale=1]{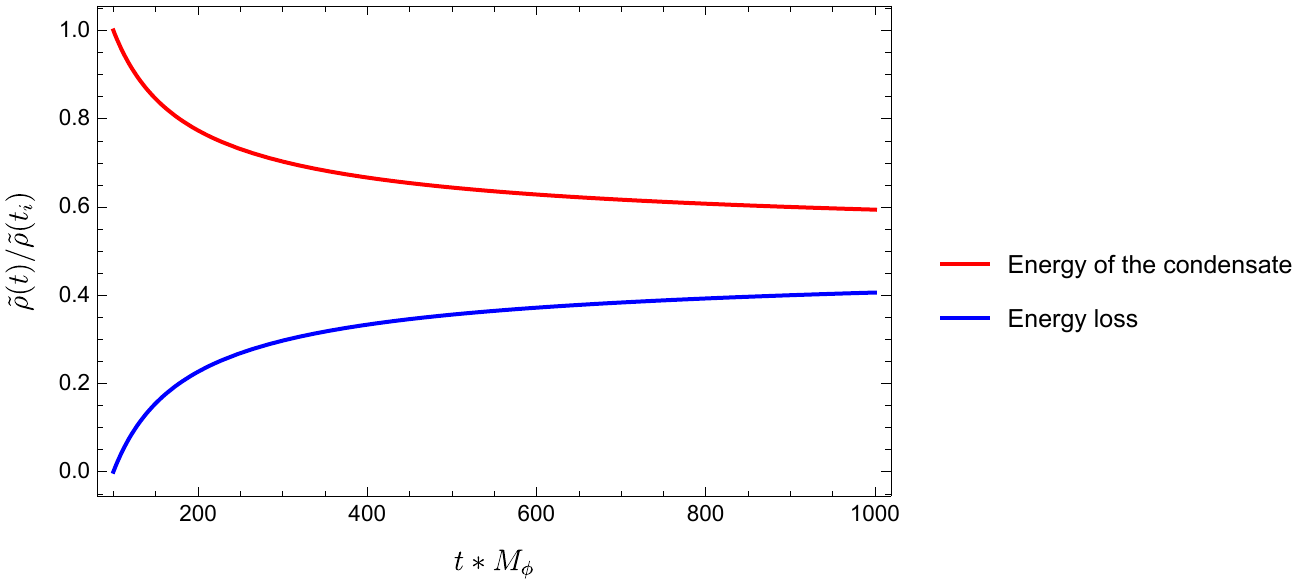}
\caption{Energy transfer from the condensate to the produced particles in a radiation-dominated universe without the self-energy correction. The evolution of the condensate comoving energy density and of the energy loss are drawn in the red and blue lines, respectively. Compared with Fig.~\ref{comovingenergy_h0_gamma0_ti=100}, the energy transfer from the condensate to the produced particles in a radiation-dominated universe is inefficient in absence of the $\gamma$ channels.  } 
\label{fig:comoving energy_gamma0_ti=100}
\end{figure}

The leading approximation for the  condensate evolution is plotted in Fig.~\ref{fig:condensate-radiation universe_gamma0_ti=100(new)}, which shows that the damping of the oscillations is dominated by the expansion of the universe. This observation suggests that reheating in this situation may not be complete. Indeed, the comoving energy density of the condensate decreases as   
\begin{align}\label{eq:comoving_rho_radiation_gamma=0}
   a^3 \rho_{\varphi} \propto \frac{1}{c_0-\frac{2\sigma}{\sqrt{t}}}\,, 
\end{align}
which approaches $1/c_0$ as $t\to \infty$\,.  Different from the case of a complete decay in a static universe (cf. Eq.~\eqref{eq:rho static universe_0}),  Eq.~\eqref{eq:comoving_rho_radiation_gamma=0} implies that, due to the expansion of the universe, the condensate will never decay completely if the $\gamma$ channels are absent (see Fig.~\ref{fig:comoving energy_gamma0_ti=100} and compare it with Fig.~\ref{comovingenergy_h0_gamma0_ti=100}). 
The remnant energy stored in the condensate depends highly on the initial conditions. 
For $t\gg t_i$, one has
\begin{align}
   \rho_{\varphi}(t) \approx \frac{t_i ^{3/2}\,\rho_{\varphi}(t_i)}{t ^{3/2}} \left( 1-\frac{4\sigma \, t_i\, \rho_{\varphi}(t_i)}{M^2_{\phi}+4\sigma \, t_i\, \rho_{\varphi}(t_i)}\right)\,.
\end{align}

\subsubsection{Non-negligible \texorpdfstring{$\gamma$}{TEXT}}

Now let us consider the full condensate equation of motion~\eqref{eq:condensate eom2}. For $\zeta =1/2$, the solution for the amplitude $A(t)$ in a radiation-dominated universe is given by
\begin{align}
\label{eq:sol-A-c=1/2}
A(t)=\frac{\e^{-\gamma t  }}{2^{1/4} \, t^{3/4} \sqrt{-\sigma  \sqrt{\gamma } \,\Gamma \left(-\frac{1}{2},2  \gamma t  \right)+c_0}}\,,
\end{align}
from which we get the solution for $f(t)$,
\begin{align}
\label{eq:sol-f-c=1/2}
f(t)=f_0+\mu  t+ \frac{ (\lambda_{\phi} +16 \alpha  M_{\phi}) \ln \left[-\sigma  \sqrt{\gamma } \,\Gamma \left(-\frac{1}{2},2  \gamma t \right)+c_0\right]}{16 M_{\phi} \sigma}\,.
\end{align}
At the leading approximation we thus have
\begin{align}
\label{eq:sol-varphi-c=1/2}
\varphi (t)\approx & \frac{\e^{-\gamma t  }}{2^{1/4} \, t^{3/4} \sqrt{-\sigma  \sqrt{\gamma } \,\Gamma \left(-\frac{1}{2},2  \gamma t  \right)+c_0}} \notag \\
&\times \cos \left\{ f_0+(M_{\phi}+\mu)  t+ \frac{ (\lambda_{\phi} +16 \alpha  M_{\phi}) \ln \left[-\sigma  \sqrt{\gamma } \,\Gamma \left(-\frac{1}{2},2  \gamma t \right)+c_0\right]}{16 M_{\phi} \sigma  }\right\}\,.
\end{align}

\begin{figure}[ht]
\centering
\includegraphics[scale=1]{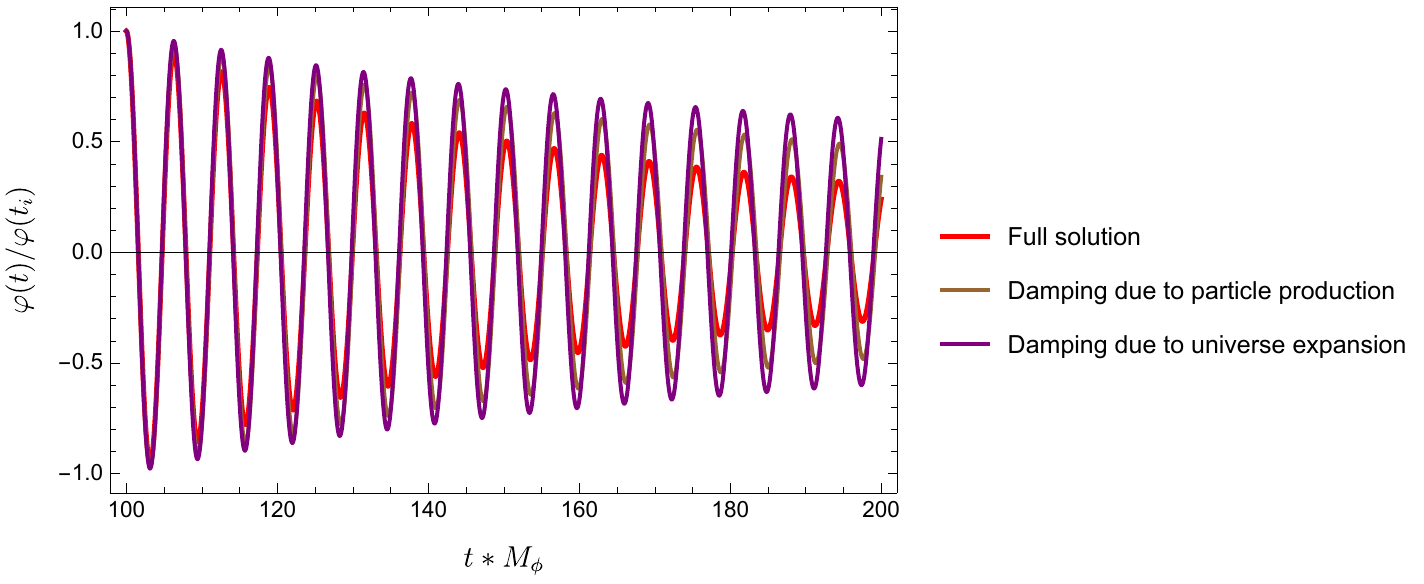}
\caption{Solutions of the condensate Eq.~\eqref{eq:condensate eom2}. Analytic approximation in a radiation-dominated universe, i.e., Eq.~\eqref{eq:sol-varphi-c=1/2},  is drawn in the red line with $\sigma=-\alpha =0.01 /M_{\varphi}$, $\mu=0.001 M_{\phi}$, $\gamma =0.005 M_{\varphi}$, and $\lambda_{\phi}=0.05$. Analytic approximation in a static universe, given by Eqs.~\eqref{eq:sol varphi_static universe1} and~\eqref{eq:sol varphi_static universe2}, is drawn in the brown line with the same parameters. Analytic approximation without interactions in a radiation-dominated universe, i.e., Eq.~\eqref{eq:varphi radiation no interation}, is drawn in the purple line.}
\label{fig:condensate-radiation universe-ti=10}
\end{figure}

\begin{figure}[ht]
\centering
\includegraphics[scale=1]{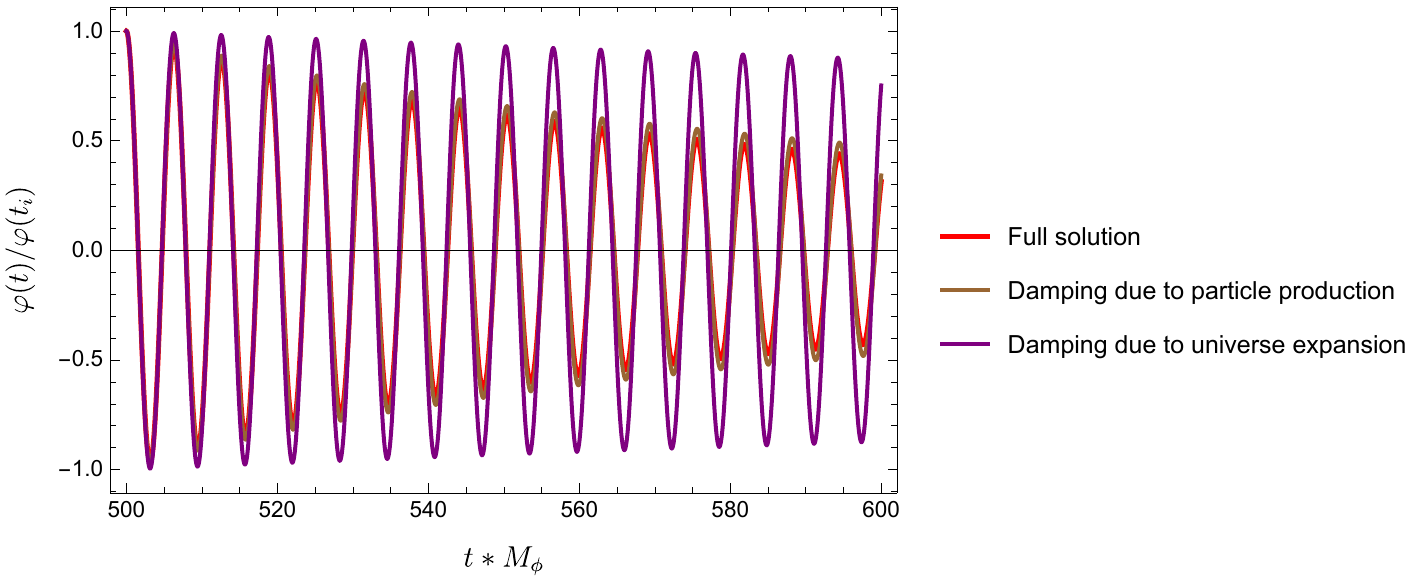}
\caption{Same as Fig.~\ref{fig:condensate-radiation universe-ti=10} but now with the initial time $t_i M_{\phi} =500$. }
\label{fig:condensate-radiation universe-ti=100}
\end{figure}

The solution~\eqref{eq:sol-varphi-c=1/2} with different initial times are drawn in the red lines in Figs.~\ref{fig:condensate-radiation universe-ti=10} and~\ref{fig:condensate-radiation universe-ti=100}. 
The earlier the condensate starts oscillating, the more important role the universe expansion plays in the damping.  

With the assumption Eq.~\eqref{eq:static universe assumption_0}, the energy density of the condensate is given by 
\begin{align}\label{eq:rho radiation universe_1}
    \rho_{\varphi} (t) \approx \frac{M^2_{\phi}}{2\sqrt{2}\, t^{3/2}}\cdot \frac{\e^{-2\gamma t  }}{c_0-\sigma  \sqrt{\gamma } \,\Gamma \left(-\frac{1}{2},2  \gamma t  \right)}\,.
\end{align}
We could also obtain the following equation for the comoving energy density of the condensate
\begin{align}\label
{eq:drho_dt_rad} 
    \frac{1}{a^3 \rho_\varphi}\frac{\d (a^3 \rho_{\varphi})}{ \d t} \approx -2\gamma -\sigma [A(t)]^2 \,.
\end{align}
The RHS of Eq.~\eqref{eq:drho_dt_rad} has the same form as the RHS  of Eq.~\eqref{eq:drho_dt_static}, but with a different expression of $A(t)$.

To understand the late-time evolution of the condensate, we consider the limit $t\gg 1/\gamma$. Then 
\begin{align}
\label{eq:radiation late time A}
A(t) 
\approx \frac{t^{-3/4}}{2^{1/4} \sqrt{c_0}}\e ^{-\gamma t} \,,
\end{align}
and 
\begin{align}
f(t)\approx f_0+\mu  t+ \frac{ (\lambda_{\phi} +16 \alpha  M_{\phi}) }{16 M_{\phi} \sigma  } \ln c_0\,,
\end{align}
where we have used the approximation that
      \begin{align}
          \Gamma \Big(-\frac{1}{2},2\gamma t \Big)\stackrel{\gamma t \gg 1}{\thickapprox } \e ^{-2 \gamma t}(2 \gamma t)^{-3/2}\,.
      \end{align}
The  solution of the condensate $\varphi$ in the late-time limit is then given by 
\begin{align}
    \varphi (t) \approx \frac{t^{-3/4}}{2^{1/4} \sqrt{c_0}}\e ^{-\gamma t} \times \cos \left[ f_0+(\mu+M_{\phi})  t+ \frac{ (\lambda_{\phi} +16 \alpha  M_{\phi}) }{16 M_{\phi} \sigma  } \ln c_0\right]\,.
    \end{align}
The comoving energy density of the condensate satisfies the following equation
\begin{align}
    \frac{1}{a^3 \,\rho_{\varphi}}\frac{\d (a^3 \rho_{\varphi})}{\d t} \approx -2 \gamma\ {\rm for}\ t\gg 1/\gamma \,, 
\end{align}
which indicates that, in the late-time limit, the comoving energy density $a^3 \rho_{\varphi}$ exponentially decreases with the decay rate $2 \gamma$\,.

\begin{figure}[H]
\centering
\includegraphics[scale=1]{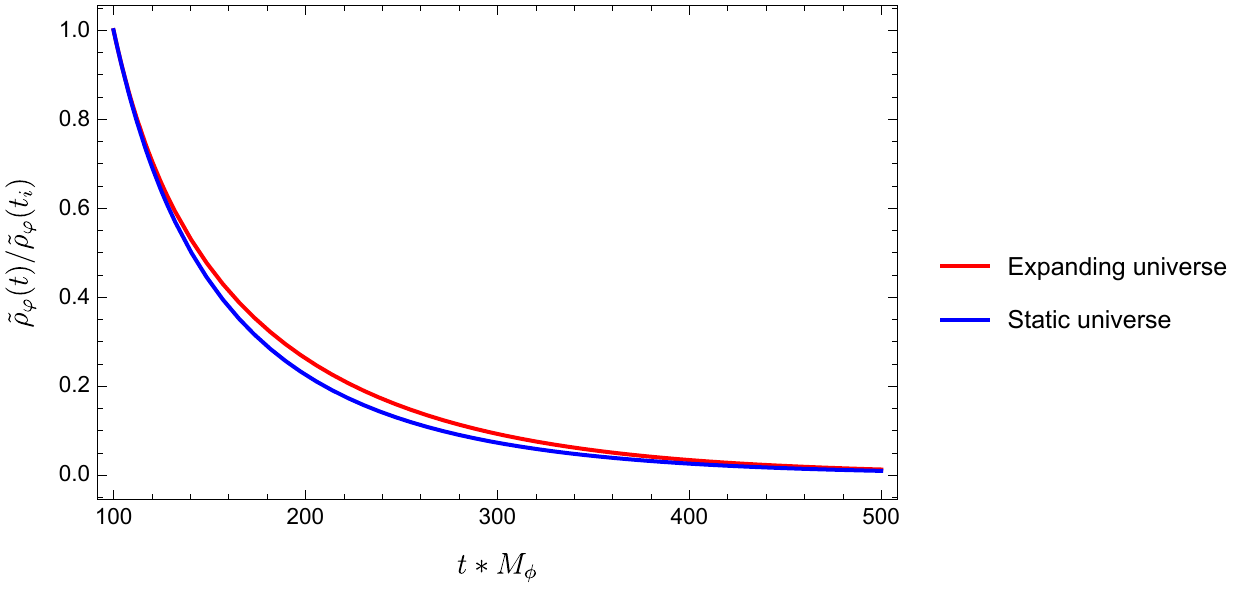}
\caption{Evolution of the (comoving) energy density of the condensate. The energy density of the condensate in a static universe, i.e.,
Eq.~\eqref{eq:rho static universe_0},
is drawn in the blue line and the comoving energy density of the condensate in a radiation-dominated universe, i.e., $a^3$ times Eq.~\eqref{eq:rho radiation universe_1}, is drawn in the red line. } 

\label{fig:energy difference}
\end{figure}

\begin{figure}[H]
\centering
\includegraphics[scale=1]{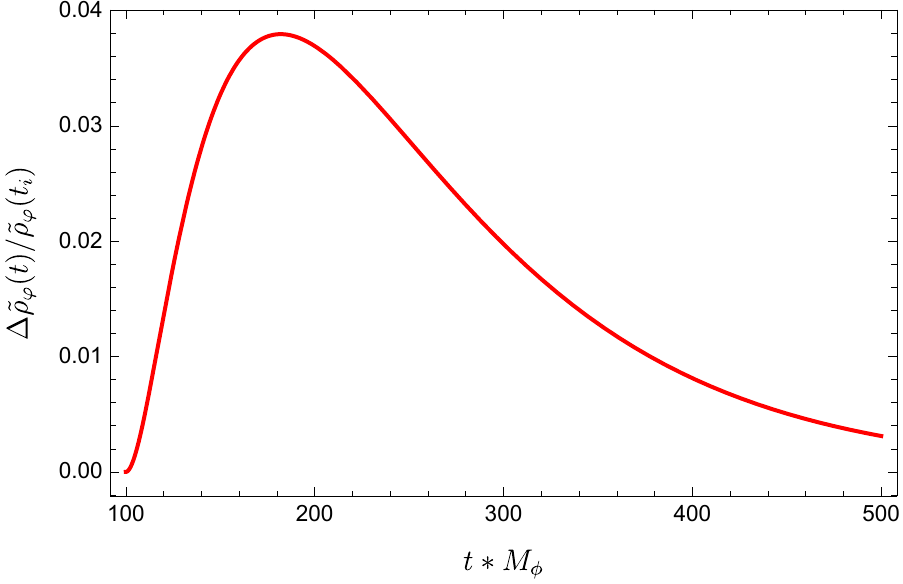}
\caption{The difference between the comoving energy density of the condensate in a radiation-dominated universe and the energy density of the condensate in a static universe.} 
\label{fig:energy difference_2.eps}
\end{figure}

Comparison between the evolution of the (comoving) energy density in a static universe 
and in a radiation-dominated universe 
is given in Figs.~\ref{fig:energy difference} and~\ref{fig:energy difference_2.eps}. From Fig.~\ref{fig:energy difference_2.eps}, we can see that the difference, denoted as $\Delta\tilde{\rho}_\varphi(t)$, grows at early times and decreases at late times. The explanation is as follows. The cubic non-local correction plays an important role in the early decay of the (comoving) energy density, cf. the $\sigma A^2$ term in  Eq.~\eqref{eq:drho_dt_static} and Eq.~\eqref{eq:drho_dt_rad}. Since the amplitude $A(t)$ decreases differently in a static and expanding universe, the (comoving) energy density also decay differently. At late times, 
their decay are both dominated by the constant $\gamma$ term, so the difference between them vanishes gradually.

\begin{figure}[H]
\centering
\includegraphics[scale=1]{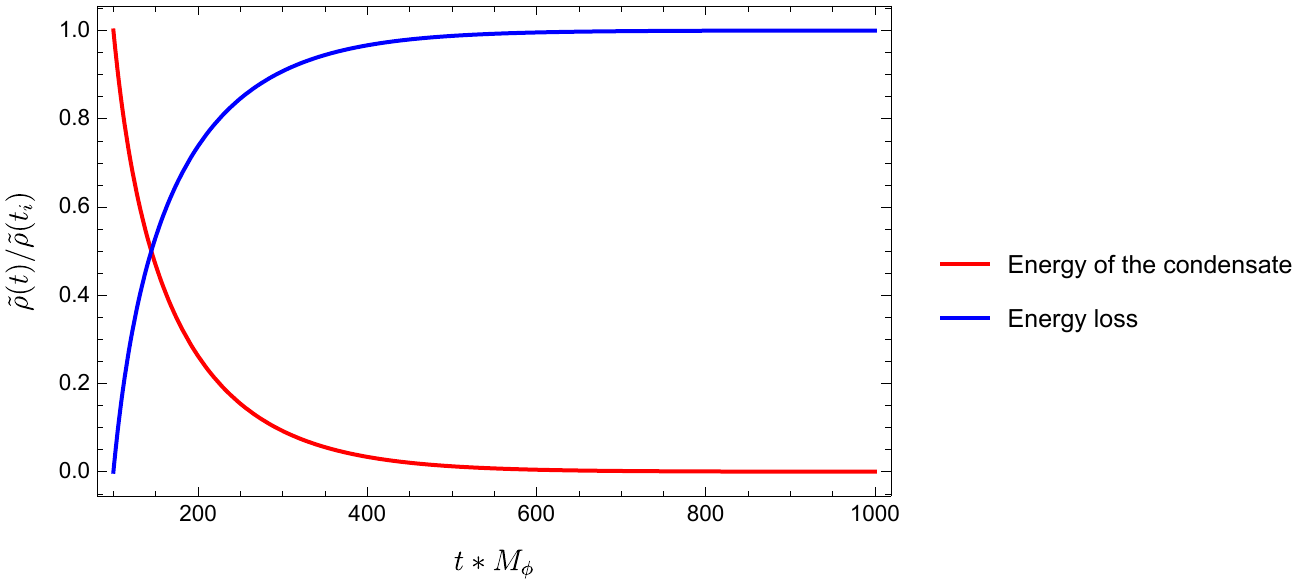}
\caption{Energy transfer from the condensate to the produced particles in a radiation-dominated universe. The evolution of the condensate comoving energy density and of the energy loss are shown in the blue and red lines, respectively. } 
\label{fig:comoving energy_ti=10}
\end{figure}

In presence of the $\gamma$ dissipation, the decay of the comoving energy density of the condensate is now complete. The energy transfer from the condensate to the produced particles in a radiation-dominated universe is shown in Fig.~\ref{fig:comoving energy_ti=10}.

So far, we have neglected the time-dependence in the microscopic quantities in deriving the solutions. In the next section, we shall discuss the effects of the time-dependence in a simpler model.

\section{Effects from the time-dependence of the microscopic quantities}
\label{app1}

In this section we take into account the time-dependence in the microscopic quantities and study how it would affect the damping of the oscillations.  However, we do not aim for a phenomenological study for reheating here and therefore we consider a few approximations to simplify the analysis. First, we ignore the backreactions from produced particles to the background spacetime and plasma temperature. Specifically, we assume a radiation-dominated universe containing the scalar condensate and additional radiation. The latter dominates the energy density, 
\begin{align}
\label{eq:constraint 1}
    \rho_{\rm rad} > \rho_{\varphi}\quad \Rightarrow\quad \frac{\pi^2}{30}g_* T^4 > \frac{1}{2}M_\phi^2||\varphi_i||^2\,,
\end{align}
where we have used the condition for the small-field regime
\begin{align}
     \frac{1}{2}M^2_\phi||\varphi_i||^2\gg \frac{\lambda_\phi}{4}||\varphi_i||^4\,,
\end{align}
such that the small-field expansion can be applied and the condensate oscillation is quasi-harmonic. For such a universe, $H\approx 1/2t$ and $T$ can be approximated by Eq.~\eqref{eq:H of T}. This approximation is valid for the very late stage of the reheating when sufficient radiation is generated from the earlier stage of the reheating. A study on the full reheating process however cannot assume fixed background spacetime and plasma. Still, the method presented here could be used to investigate whether reheating in a given model is complete or not. Second, we only consider the self-interacting $\Phi^4$ theory which is  too ideal to be a realistic reheating model.

\begin{figure}[H]
\centering
\includegraphics[scale=0.9]{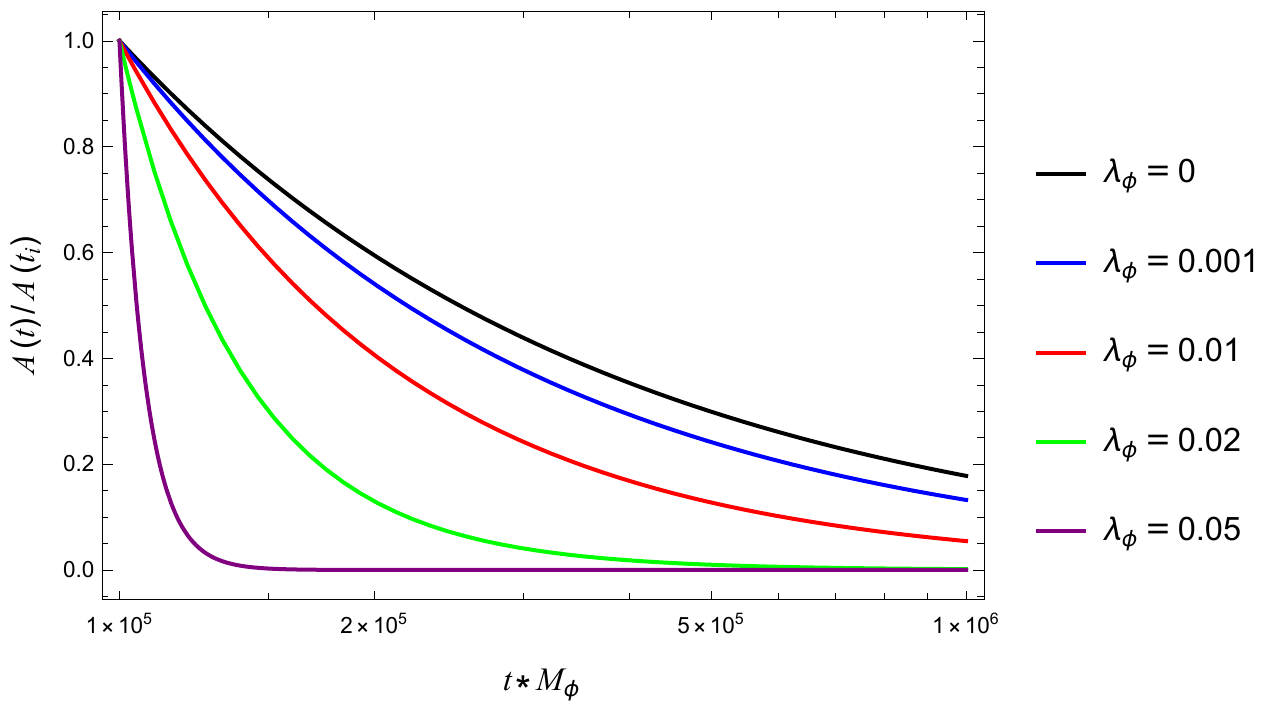}
\caption{Numerical solutions of the condensate equation~\eqref{eq:app_eomA} with $M_{\rm pl}/M_{\phi}=10^9$, $g_* =100$, and various values for $\lambda_\phi$. The initial condition $A(t _i)/M_{\phi}=1$ is set at $t _i M_{\phi} =10^5$. 
The parameters have been chosen in such a way that Eq.~\eqref{eq:constrain M approx m} is satisfied.
} 
\label{fig:appfigAtau}
\end{figure}

\begin{figure}[H]
\centering
\includegraphics[scale=0.9]{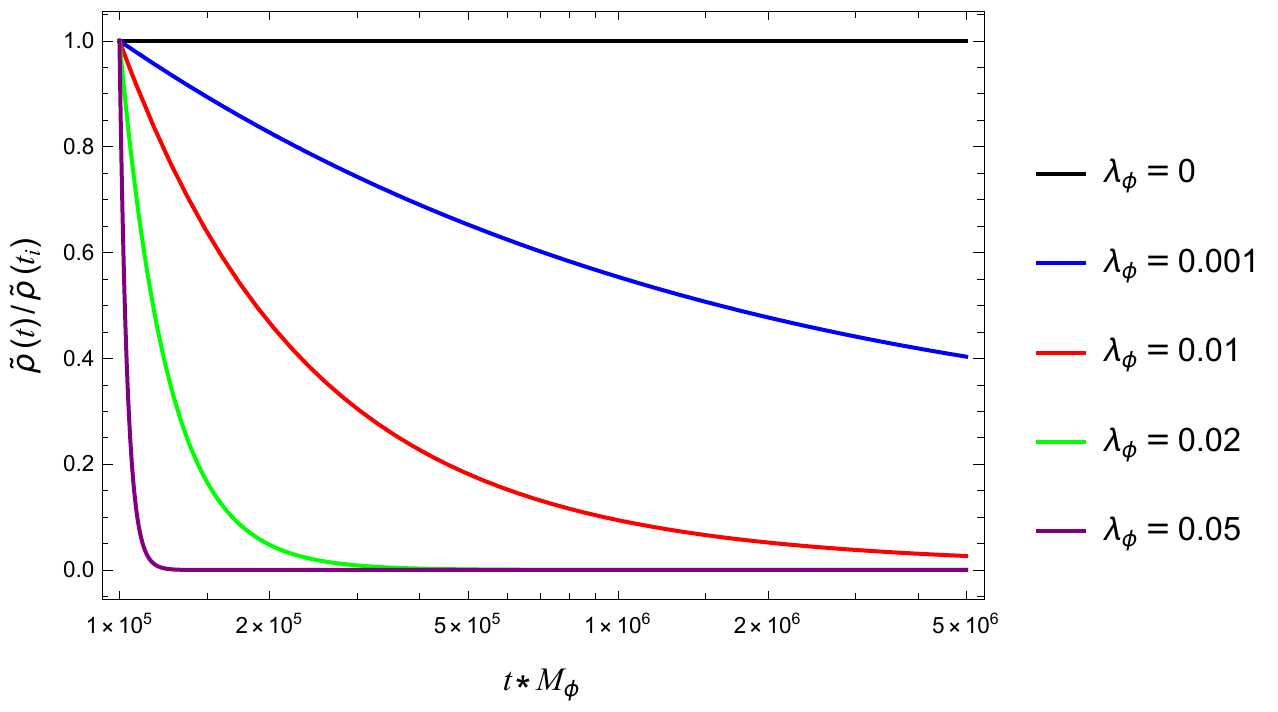}
\caption{Evolution of the condensate comoving energy density for the numerical solutions given in Fig.~\ref{fig:appfigAtau}.} 
\label{fig:appfigenergy}
\end{figure}

Since we are interested in the decay of the condensate, we shall look at only the evolution of the envelope of the oscillations.  In this simple model, $\sigma$ channels are absent, and we have
\begin{align}
   \gamma= -\frac{{\rm Im} \left[\widetilde{\pi}_{\rm R,\phi}(M_\phi)\right]}{2M_{\phi}}=\frac{\lambda_\phi^2 T^2}{256M_{\phi} \pi^3}{\rm Li}_2\left(\e^{-M_\phi/T}\right)\,.
\end{align}
Using Eq.~\eqref{eq:H of T}, one obtains 
\begin{align}
\label{eq:T(t)}
    T\approx \left(\frac{45 M^2_{\rm pl}}{4\pi^3 g_*}\right)^{1/4} \sqrt{H}=\frac{1}{2}\left(\frac{45 M^2_{\rm pl}}{\pi^3 g_*}\right)^{1/4}\frac{1}{\sqrt{t}}\,.
\end{align}
Therefore, $\gamma$ is time dependent. In the multi-scale analysis, the microscopic quantities depend on the slow time $\tau$ as $H$ does. As a result, the only modification in the multi-scale analysis is that in Eqs.~\eqref{eq:eom R_1} and~\eqref{eq:eom_A and f(tau)}, one has to write the microscopic quantities as $\gamma(\tau)$, $\mu(\tau)$ etc. The equation of motion for $A(\tau)$ reads
\begin{align}\label{eq:app_eomA}
    \frac{\d A (\tau)}{\d \tau}+\left(\gamma(\tau) +\frac{3}{2}H(\tau)\right) A(\tau) =0\,,
\end{align}
where 
\begin{align}
    \gamma (\tau)=\frac{3\lambda_{\phi}^2}{1024\, \pi^4}\cdot\sqrt{\frac{5}{\pi g_*}}\cdot\frac{M_{\rm pl}}{M_{\phi}}\cdot \frac{1}{\tau}\cdot {\rm Li}_2\left[\exp{\left(-2M_{\phi}\sqrt{\tau}\cdot\left(\frac{\pi^3 g_*}{45M_{\rm pl}^2}\right)^{1/4}\right)} \right]\,.
\end{align}
One needs to take $\tau=t$ after $A(\tau)$ is obtained. Numerical solutions of Eq.~\eqref{eq:app_eomA}  are given in Fig.~\ref{fig:appfigAtau}. The corresponding evolution of the comoving energy density for the condensate is plotted in Fig.~\ref{fig:appfigenergy}. Bigger $\lambda_{\phi}$ or $M_{\rm pl}/M_{\phi}$ makes the energy transfer from the condensate to the produced particles quicker. In choosing the parameters, one has to remember the condition that
$M_\phi\approx m_\phi$  (see the comment below Eq.~\eqref{eq:eom varphi_1_0}). Let $M^2_\phi= m^2_\phi + \Delta m^2_\phi(T)$. Since the upper bound
for $\Delta m^2_\phi(T)$ (for given $T$) is the high-temperature limit $\lambda_\phi T^2/24$, the condition
\begin{align}
    M^2_\phi\gg \frac{\lambda_\phi}{24} T^2
\end{align}
is sufficient to ensure $M^2_\phi\approx m^2_\phi$.
Using Eq.~\eqref{eq:T(t)}, The above constraint becomes
\begin{align}
\label{eq:constrain M approx m}
     \tau\cdot M_{\phi}\gg \frac{\lambda_{\phi}}{32}\cdot\frac{M_{\rm pl}}{M_{\phi}}\cdot\sqrt{\frac{5}{\pi ^3 g_*}}\approx \O (10^{-3})\, \lambda_\phi \cdot\frac{M_{\rm pl}}{M_\phi} \,,
\end{align}
where we have assumed $g_*\approx \O(100)$.

\begin{figure}[H]
\centering
\includegraphics[scale=0.9]{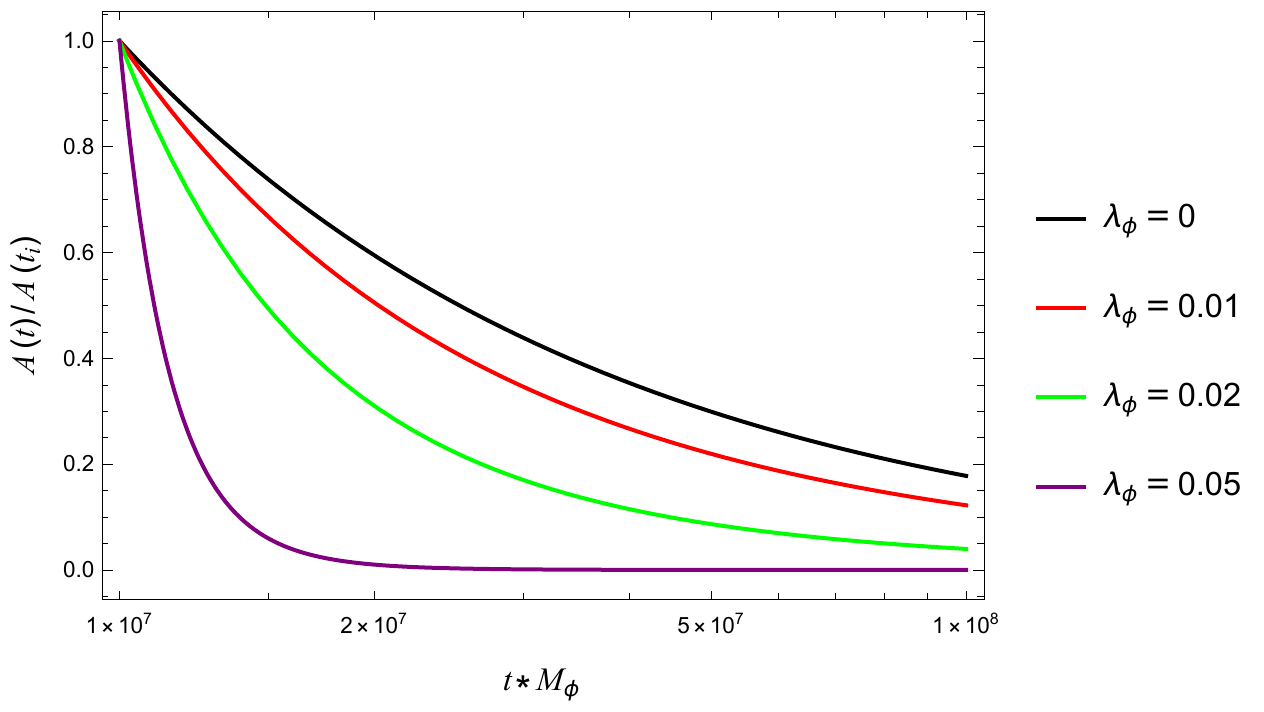}
\caption{Numerical solutions of the condensate equation~\eqref{eq:app_eomA} with $M_{\rm pl}/M_{\phi}=10^9$ and $g_* =100$. The initial condition $A(t _i)/M_{\phi}=1$ is set at $t _i M_{\phi} =10^7$ such that the initial temperature $T_i\approx M_{\phi}$\,.  A larger $\lambda_{\phi}$ corresponds to a larger $\Omega$ and the energy transfer from the condensate to the produced particles is more efficient.} 
\label{fig:appfigAtau3}
\end{figure}

\begin{figure}[H]
\centering
\includegraphics[scale=0.9]{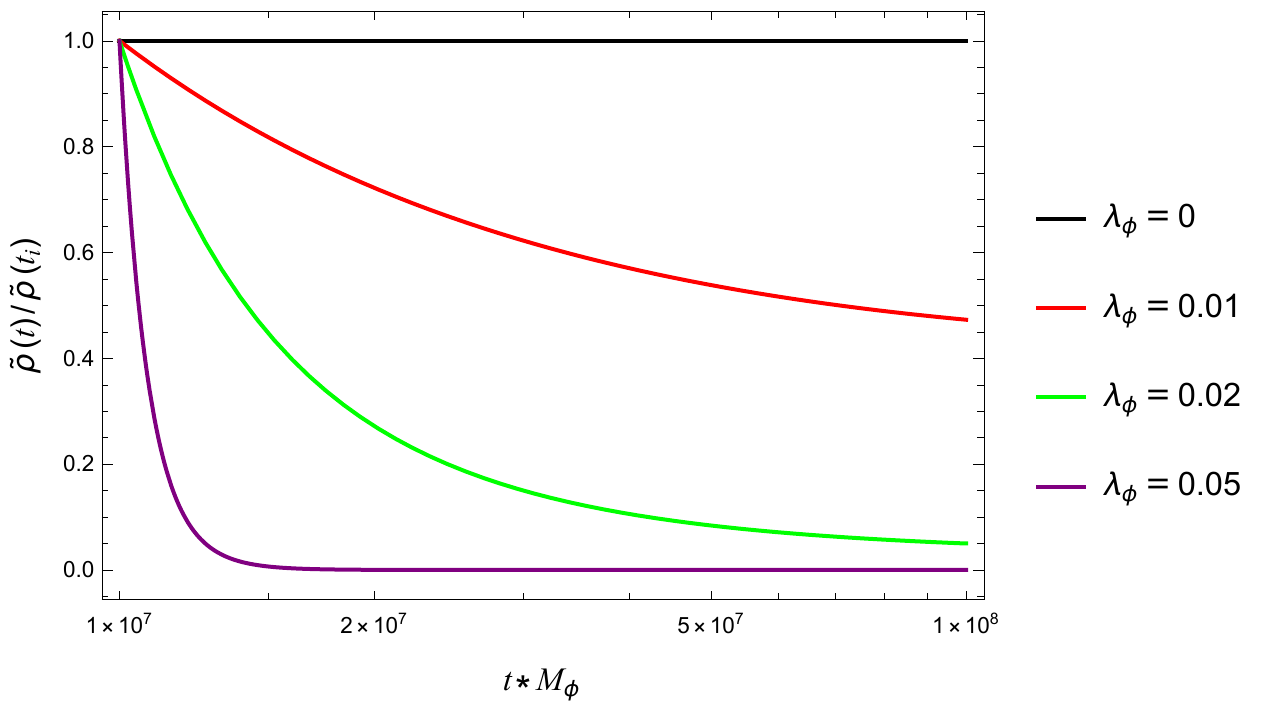}
\caption{Evolution of the condensate comoving energy density for the numerical solutions given in Fig.~\ref{fig:appfigAtau3}.} 
\label{fig:appfigenergy3}
\end{figure}

To better understand the numerical solutions discussed above, let us study the analytical solution of Eq.~\eqref{eq:app_eomA} in the high-temperature limit ($T\gg M_{\phi}$), which gives the constraint
\begin{align}
     \left(\frac{\lambda_{\phi}}{32}\cdot\frac{M_{\rm pl}}{M_{\phi}}\cdot\sqrt{\frac{5}{\pi ^3 g_*}} \right) \ll\tau\cdot M_{\phi}\ll \frac{3}{4} \cdot\frac{M_{\rm pl}}{M_{\phi}}\cdot \sqrt{\frac{5}{\pi^3 g_*}}\,.
\end{align}
where the second inequality is due to $T\gg M_{\phi}$.
If we take, e.g., $g_{*}=100$, $M_{\rm pl}/M_{\phi} =10^9$ and $\lambda_{\phi} =0.01$, we would have the constraint $10^{4} \ll \tau \cdot M_{\phi}  \ll 3 \times 10^7$. Applying the high-temperature limit of the dilogarithm function, Eq.~\eqref{eq:Li2-highT}, we obtain 
\begin{align}
\label{eq:gamma-Omega-H}
    \gamma \simeq \Omega \cdot H\,,
\end{align}
where the dimensionless coefficient $\Omega$ is given by\begin{align}\label{eq:Omega_def}
  \Omega =\frac{M_{\rm pl}}{M_{\phi}}\cdot \frac{\lambda_\phi^{2}}{1024 \pi ^2} \cdot \sqrt{\frac{5}{\pi g_{*}}}\,.
\end{align}
Substituting Eq.~\eqref{eq:gamma-Omega-H} and $H(\tau)=1/(2\tau)$ into Eq.~\eqref{eq:app_eomA}, we obtain the solution
\begin{align}
       A(\tau)=\frac{c_0}{\tau ^{3/4+\Omega/2}}\,,
\end{align}
where $c_0$ is a constant of integration. Note that the power-law damping $\tau ^{\Omega/2}$ is caused by the $\gamma$ channel $\varphi\phi\rightarrow\phi\phi$ and the power-law damping $\tau ^{3/4}$ is caused by the expansion of the universe. If $\Omega\ll 3/2$, the damping of the condensate is dominated by the expansion of the universe. So, to have an efficient energy transfer in this simple situation, we must  have $\Omega > 3/2$.  Bigger $\Omega$ should make the energy transfer from the condensate to the produced particles quicker. This observation agrees with our numerical results given in Figs.~\ref{fig:appfigAtau} and \ref{fig:appfigenergy} where the red lines correspond $\Omega \approx 1$. In  Figs.~\ref{fig:appfigAtau3} and \ref{fig:appfigenergy3}, we show the numerical solutions in the regime $T\lsim M_\phi$, and the conclusion is also valid for that regime.

In the last section, we concluded that the $\gamma$ channel is necessary to ensure a complete decay of the condensate. This conclusion does not change if we take into account the time-dependence in $\gamma$, but the latter would enforce further constraint on the coupling constants by requiring $\Omega\gg 1$. For the simple model analysed here and in the high-temperature regime, it is
\begin{align}
    \frac{M_{\rm pl}}{M_{\phi}}\cdot \lambda_\phi^{2}\gg1024 \pi ^2 \cdot \sqrt{\frac{\pi g_{*}}{5}}\,.
\end{align}
When the time-dependence in $\gamma$ is neglected, the term $\gamma A(\tau)$ in Eq.~\eqref{eq:app_eomA} always dominates over the Hubble term at late times because $H(\tau)$ keeps decreasing. Therefore, the decay of the condensate is complete. However, when the time-dependence in $\gamma$ is taken into account, the efficiency of the decay through the $\gamma$ channel also decreases with time. And if $\lambda_\phi$ is too small, the $\gamma(\tau)A(\tau)$ term would be smaller than the Hubble term at late times such that the decay never completes. Going beyond the high-temperature regime, the constraint should be more strict because the channel would be closed (or exponentially suppressed) when $T\ll M_\phi$.

\section{Conclusions}
\label{sec:Conc}

In this paper we studied the damping of scalar condensate oscillations in a spatially flat FLRW universe in the mildly non-linear regime, generalizing the work~\cite{Ai:2021gtg}. We use non-equilibrium quantum field theory which naturally accommodates quantum statistical effects of the plasma as well as the backreaction effects from particle production. Even though the framework is quite general, we used a $Z_2$-symmetric two-scalar model with quartic interactions as a concrete example. The starting point of our analysis is the non-local equation of motion~\eqref{eq:condensate eom2}. The information about particular interactions is encoded in the integral kernels $\pi_{\rm R}(t-t')$ and $v_{\rm R} (t-t')$, which are closely related to the retarded self-energies and proper four-vertex functions in a given particle physics model. We applied the multi-scale analysis to solve the non-local equation~\eqref{eq:condensate eom2}, and were able to find leading-order analytic approximations for the solution in terms of the Fourier-transformed retarded self-energy and proper four-vertex function evaluated at particular energies that are related to the oscillation frequency. We carry out the analysis in static, radiation-dominated, and matter-dominated universes (see the Appendix). The solutions take the general form $\varphi(t)\approx A(t)\times \cos[f(t)+M_\phi t]$ and are concluded in Table~\ref{tab:label1}. 

\begin{table}[ht]
    \centering
    \begin{tabular}{|c|c|c|}
    \hline
             & $A(t)$ & $f(t)$ \\
    \hline         
    static & $\frac{A_0 \e^{-\gamma t}}{\sqrt{1+\frac{\sigma A_0^2}{2\gamma}(1-\e^{-2\gamma t})}}$ & $f_0+\mu t +\frac{ (\lambda_{\phi} +16 \alpha  M_{\phi}) \ln \left[1+\frac{\sigma A^2 _0}{2 \gamma}(1-\e^{-2\gamma t})\right]}{16 M_{\phi} \sigma}$ \\
    \hline
    radiation-dominated & $\frac{\e^{-\gamma t  }}{2^{1/4} \, t^{3/4} \sqrt{-\sigma  \sqrt{\gamma } \,\Gamma \left(-\frac{1}{2},2  \gamma t  \right)+c_0}}$ & $f_0+\mu  t+ \frac{ (\lambda_{\phi} +16 \alpha  M_{\phi}) \ln \left[-\sigma  \sqrt{\gamma } \,\Gamma \left(-\frac{1}{2},2  \gamma t \right)+c_0\right]}{16 M_{\phi} \sigma  }$ \\
    \hline
    matter-dominated & $\frac{ \e^{- \gamma t}}{t\,\sqrt{  c_0-2 \sigma \, \gamma \, \Gamma (-1,2 \gamma t)}}$ & $f_0+\mu  t+\frac{(\lambda_{\phi} +16 \,\alpha \, M_{\phi}) \ln \left[c_0-2 \, \sigma \, \gamma\, \Gamma (-1,2 \gamma t  )\right]}{16 M_{\phi} \sigma}$ \\
    \hline
    \end{tabular}
    \caption{Solutions for the condensate evolution in a static, radiation-dominated, and matter-dominated universe. The constants $A_0$, $c_0$, and $f_0$ can be determined by the initial conditions for $\varphi$ and $\dot{\varphi}$.}
    \label{tab:label1}
\end{table}

In obtaining these solutions, we have assumed that the microscopic quantities are time independent. This is not consistent when the temperature is evolving. To fully take into account the time-dependence in the microscopic quantities, one needs to know their closed expressions in terms of the couplings and temperature. Taking into account the time-dependence in the microscopic quantities is much more tricky. In Sec.~\ref{app1}, we show in a simple situation how these effects may affect the conclusions. A dedicate study of perturbative reheating which not only takes into account the time-dependence in the microscopic quantities but also solves the coupled equations of motion for the inflaton, radiation and spacetime will be given in Ref.~\cite{AL}. Aside from the above issue, we have also employed two other assumptions in this paper. First, we have used the small-field expansion which assumes that the $\varphi$-dependent mass terms are smaller than the $\varphi$-independent mass terms of the same field. Second, we have assumed that $M^2_\phi\approx  m^2_\phi$. The first assumption also implies that the inflaton potential is dominated by the quadratic term such that the oscillation is quasi-harmonic.

As in the case of flat spacetime, in an expanding universe the dissipation is due to particle production from the decaying oscillating condensate. The decay channels can be classified into two classes. The first class is characterized by a non-vanishing imaginary part of the retarded self-energy, the $\gamma$ defined in Eq.~\eqref{eq:important_quantities}, and contains the processes given in Eq.~\eqref{eq:decaychannel-one}. We call them the $\gamma$ channels. These channels give rise to Landau damping with one condensate quantum and manifest them in the equation of motion through the linear non-local term associated with $\pi_{\rm R}(t-t')$. The second class is characterized by a non-vanishing imaginary part of the retarded four-vertex function, the $\sigma$ defined in Eq.~\eqref{eq:important_quantities}, and contains the process given in Eq.~\eqref{eq:decaychannel-two}. We call it the $\sigma$ channel. The $\sigma$ channel can also happen at zero temperature. They are encoded in the cubic non-local term in the equation of motion.

An important observation made in this work is that the $\gamma$ channels are necessary to ensure a complete decay of the condensate (this conclusion can be generalized to the case of a temperature-dependent $\gamma$, see Sec.~\ref{app1}). The evolution of the condensate energy can be generally described by the following equation
\begin{align}
    \frac{1}{a^3\rho_\varphi}\frac{\d (a^3\rho_\varphi)}{\d t}\approx -2\gamma-\sigma [A(t)]^2\,.
\end{align}
The $\gamma$ term is constant and thus induces exponential damping for the comoving energy density $a^3\rho_\varphi$, while the $\sigma$ term falls off with the amplitude squared and induces power-law damping. For early times when the oscillation amplitude is still large, the comoving energy density first experiences a power-law damping behavior and at later times an exponential damping behavior. In flat spacetime, the decay of the condensate is complete even if $\gamma$ is negligible, i.e., $\gamma=0$. The conclusion is completely different in an expanding universe. When the $\gamma$ dissipation is absent, the energy density of the condensate in an expanding universe satisfies
\begin{align}
   \frac{1}{\rho_{\varphi}(t)}\, \frac{\d \rho_{\varphi}(t)}{d t} \approx -3H(t)-\sigma [A(t)]^2\,.
\end{align}
In an expanding universe, $[A(t)]^2$ decreases faster than the Hubble constant (for a radiation-dominated universe, $[A(t)]^2$ decreases as $t^{-3/2}$ and for a matter-dominated universe it decreases as $t^{-2}$). The decrease of the energy density at late times is thus dominantly caused by the expansion of the universe and the energy transfer from the condensate to the produced particles never completes. This can also be very clearly seen from our explicit solutions whose behaviors have been extensively shown in various plots throughout the paper. When taking into account the time-dependence in the microscopic quantities, say $\gamma$ and $\sigma$, one would have further constraints to ensure a complete decay of the condensate. This is because $\gamma(t)$ in general decreases with time. As such there is a new competition between $H(t)$ and $\gamma(t)$, in contrast to the case when $\gamma$ is regarded as  constant and always dominates over $H(t)$ at sufficiently late times.

\section*{Acknowledgments}

WYA thanks Gilles Buldgen, Marco Drewes, Dra\v{z}en Glavan, Jan Hajer for many discussions on non-equilibrium quantum field theory. The work of ZLW is supported by the Natural Science Research Project of Colleges and Universities in JiangSu Province (21KJB140001) and Natural Science Foundation of Jiangsu Province (BK20220642).

\begin{appendix}
\renewcommand{\theequation}{\Alph{section}\arabic{equation}}

\section{Matter-dominated universe}
In this Appendix, we include a discussion on the solution of the condensate in a  matter-dominated universe ($\zeta=2/3$). In such a universe and for processes whenever the Hubble expansion is relevant, the temperature is typically negligibly small. In this case, $\gamma$ is vanishing and Landau damping through the processes in Eq.~\eqref{eq:decaychannel-two} is absent. However, for theoretical interest, we still discuss the situation with non-vanishing $\gamma$ below.  

\subsection{No interactions}
     
First consider the evolution of a free massive scalar field in the background of a matter-dominated universe. This could be realized by setting $\gamma =\sigma=\lambda_{\phi}=\mu =\alpha=0$ in Eq.~\eqref{eq:sol_varphi_0} and taking $\zeta=2/3$. Then we have 
\begin{subequations}
\begin{align}
\label{eq:sol_A_0 c=2/3}
A(t)=&\frac{1}{ \sqrt{ c_0}\,t}\,,\\
f(t)=&f_0\,,
\end{align}
\end{subequations} 
where $f_0$ is a constant of integration. In this case, the leading approximation for the solution of $\varphi (t)$ is given by
\begin{align}
\label{eq:sol_varphi_0 c=2/3}
\varphi(t)\approx \frac{1}{ \sqrt{ c_0  }\,t}\times \cos\left(f_0+M_{\phi}t\right)\,,
\end{align}
which agrees with the result given in Ref.~\cite{Mukhanov:2005ca}.\footnote{In our calculation, we have treated the spacetime as a fixed background. 
However, a universe with only an oscillating scalar background field with a quadratic potential does behave like a matter-dominated universe.}

The amplitude of the oscillations, given in  Eq.~\eqref{eq:sol_A_0 c=2/3}, falls off as $[a(t)]^{-3/2}$. As in the case of a radiation-dominated universe the total energy of the condensate, proportional to $a^3\rho_\varphi$, is conserved.

\subsection{Negligible \texorpdfstring{$\gamma$}{TEXT}}

\begin{figure}[ht]
\centering
\includegraphics[scale=1]{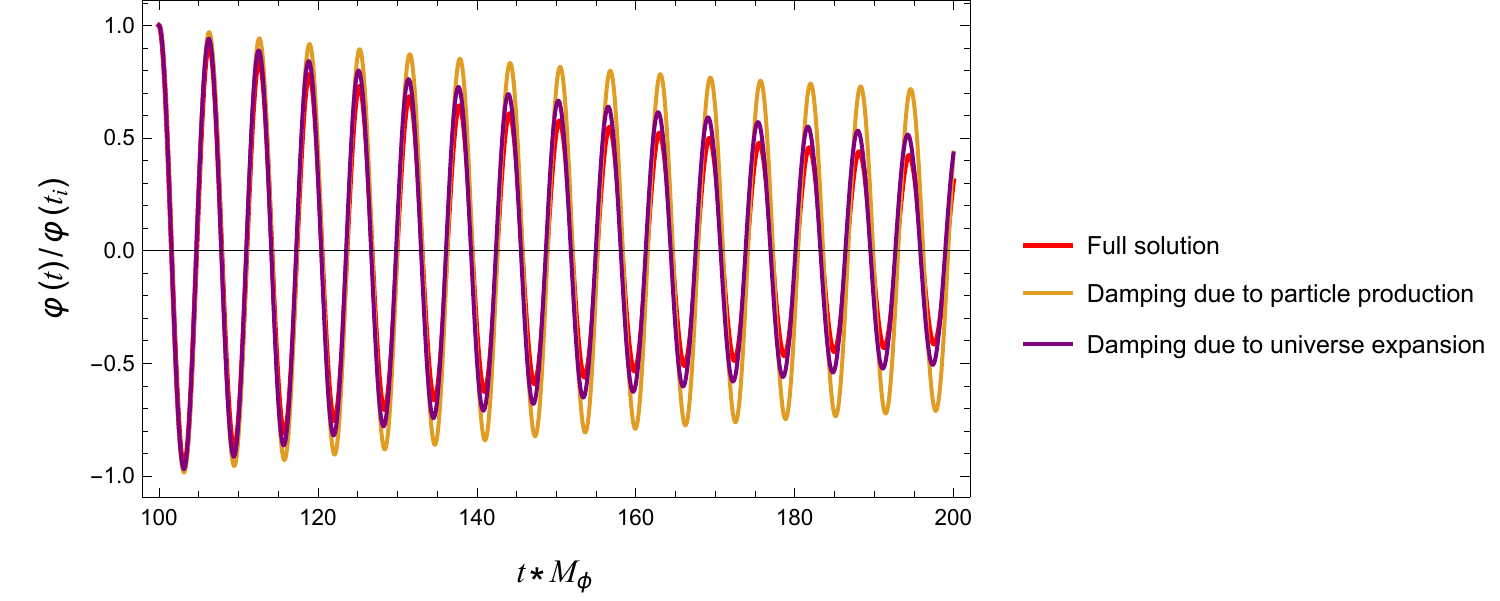}
\caption{Solutions of the condensate equation \eqref{eq:condensate eom2} with the imaginary part of the self-energy neglected $\gamma =0 $. Analytic approximation in a matter-dominated universe, i.e., Eq.~\eqref{eq:sol-varphi-gamma=0_c=2/3},  is drawn in the red line with $\sigma=-\alpha=0.01/ M_{\varphi}$, $\lambda_{\phi}=0.05$, $\gamma=0$, and $\mu =0.001 M_{\phi}$. Analytic approximation in a static universe with the same parameters, given by Eqs.~\eqref{eq:sol varphi_static universe1_gamma0}, is drawn in the brown line. Analytic approximation without interactions in a matter-dominated universe, i.e., Eq.~\eqref{eq:sol_varphi_0 c=2/3}, is   drawn in the purple line.}
\label{fig:sol_matter_gamma0_ti=10}
\end{figure}

Now we consider the case where the imaginary part of the self-energy is negligible. Then, we find
\begin{subequations}
\begin{align}\label{eq:sol-varphi-gamma=0_c=2/3}
A(t)&= \frac{1}{\sqrt{-\sigma  \,t+c_0\, t^2}}\,,\\
f(t)&=f_0 +\mu t-\frac{(\lambda_{\phi} +16 \alpha \, M_{\phi}) \left[\ln (t)-\ln (c_0\, t-\sigma )\right]}{16 \,M_{\phi} \,\sigma }\,,
\end{align}
\end{subequations}
where $c_0$ and $f_0$ are constants of integration. The leading approximation for the solution of condensate is plotted in the red curve in Fig.~\ref{fig:sol_matter_gamma0_ti=10}. The damping of the oscillation is dominated by the expansion of the universe.

\begin{figure}[ht]
\centering
\includegraphics[scale=1]{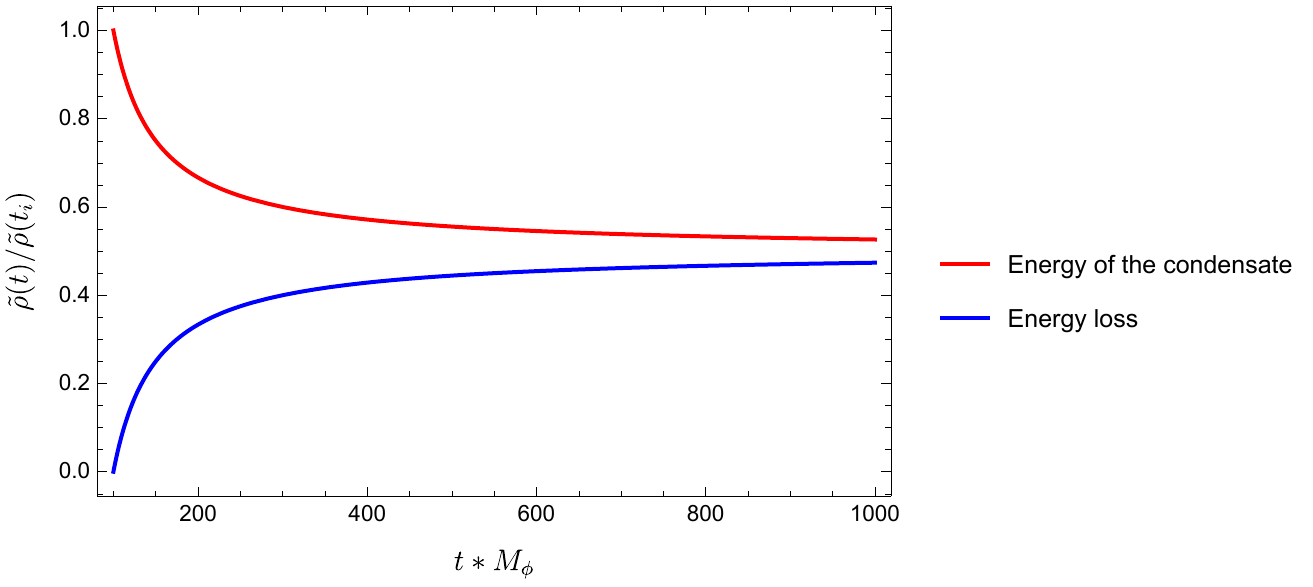}
\caption{Energy transfer from the condensate to the produced particles in a matter-dominated universe without the self-energy correction. The evolution of the condensate comoving energy density and of the energy loss are drawn in the red and blue lines, respectively. Comparing with Fig.~\ref{comovingenergy_h0_gamma0_ti=100}, the energy transfer from the condensate to the produced particles in a matter-dominated universe is inefficient if the $\gamma$ dissipation is absent.} 
\label{fig:energy_matter_gamma0_ti=10}
\end{figure}

The comoving energy density of the condensate decreases as   
\begin{align}\label{eq:comoving_rho_matter_gamma=0}
   a^3 \rho_{\varphi} \propto \frac{1}{c_0-\frac{\sigma}{t}}\,, 
\end{align}
which approaches $1/c_0$ as $t\to \infty$\,. The comoving energy density of the condensate will never decay completely, see Fig.~\ref{fig:energy_matter_gamma0_ti=10}. 
For $t\gg t_i$,
\begin{align}
   \rho_{\varphi}(t) \approx \frac{t_i ^{2}\,\rho_{\varphi}(t_i)}{t ^{2}} \left( 1-\frac{2\sigma \, t_i\, \rho_{\varphi}(t_i)}{M^2_{\phi}+2\sigma \, t_i\, \rho_{\varphi}(t_i)}\right)\,.
\end{align}

\subsection{Non-negligible \texorpdfstring{$\gamma$}{TEXT}}

\begin{figure}[H]
\centering
\includegraphics[scale=1]{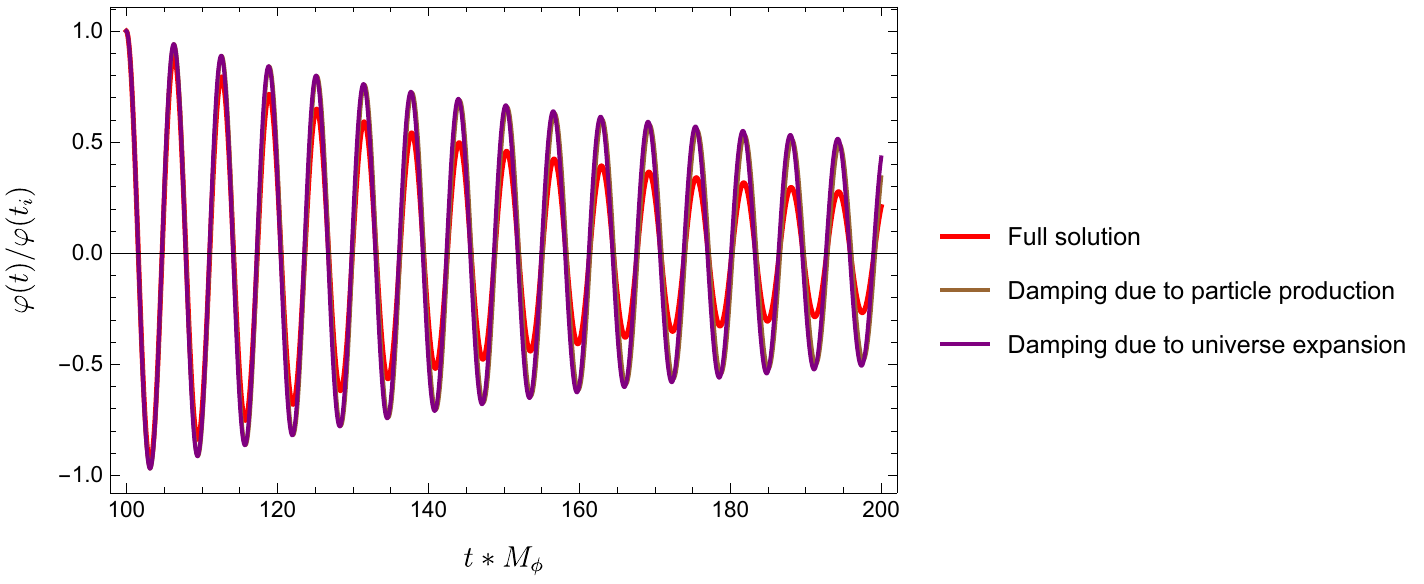}
\caption{Solutions of the condensate equation \eqref{eq:condensate eom2}. Analytic approximation in a matter-dominated universe, i.e., Eq.~\eqref{eq:sol_matter_full_0},  is drawn in the red line with $\sigma=-\alpha =0.01/ M_{\phi}$, $\mu =0.01 M_{\phi}$, $\gamma =0.005 M_{\phi}$, $\lambda_{\phi}=0.05$. Analytic approximation in a static universe with the same parameters, given by Eqs.~\eqref{eq:sol varphi_static universe1} and~\eqref{eq:sol varphi_static universe2}, is drawn in the brown line. Analytic approximate solution without interactions in a matter-dominated universe, i.e., Eq.~\eqref{eq:sol_varphi_0 c=2/3}, is drawn in the purple line.}
\label{fig:sol_matter_full_ti=10}
\end{figure}

\begin{figure}[H]
\centering
\includegraphics[scale=1]{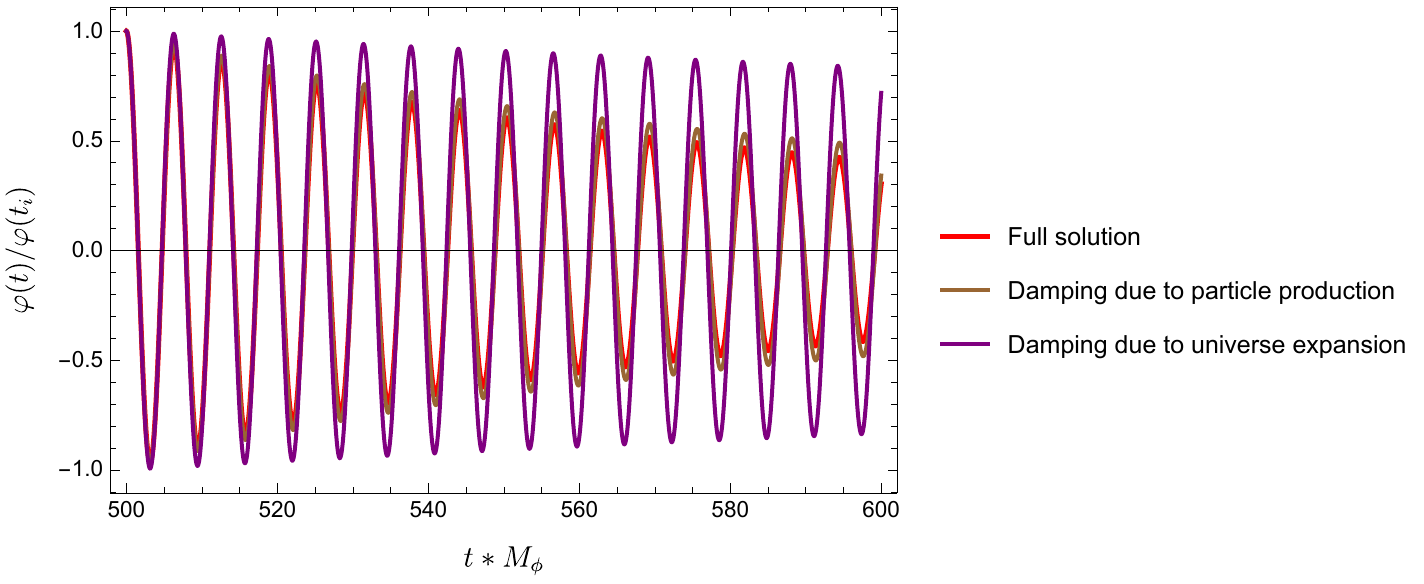}
\caption{Same as Fig.~\ref{fig:sol_matter_full_ti=10} but now with initial time $t_i M_{\phi}=500$.}
\label{fig:sol_matter_full_ti=100}
\end{figure}

Finally, let us consider the full condensate equation of motion~\eqref{eq:condensate eom2} in a matter-dominated universe. We find
\begin{subequations}
\begin{align}\label{eq:matter_full_A}
   A(t)&= \frac{ \e^{- \gamma t}}{t\,\sqrt{  c_0-2 \sigma \, \gamma \, \Gamma (-1,2 \gamma t)}}\,,\\
   f(t)&= f_0+\mu  t+\frac{(\lambda_{\phi} +16 \,\alpha \, M_{\phi}) \ln \left[c_0-2 \, \sigma \, \gamma\, \Gamma (-1,2 \gamma t  )\right]}{16 M_{\phi} \sigma}\,,
\end{align}
\end{subequations}
where $c_0$ and $f_0$ are constants of integration. Then, the leading approximation for the solution of Eq.~\eqref{eq:condensate eom2} in a matter-dominated universe reads 
\begin{align}
\label{eq:sol_matter_full_0}
   \varphi \approx  \frac{ \e^{- \gamma t}}{t\,\sqrt{  c_0-2 \sigma \, \gamma  \,\Gamma (-1,2 \gamma t)}}\, \cos\left\{ f_0+(M_{\phi}+\mu)  t+\frac{(\lambda_{\phi} +16 \alpha  M_{\phi}) \ln \left[c_0\,-2\sigma\gamma \Gamma (-1,2 \gamma t  )\right]}{16 \,M_{\phi} \,\sigma } \right\}\,.
\end{align}
The solution Eq.~\eqref{eq:sol_matter_full_0} with different initial times are drawn in the red curves in Fig.~\ref{fig:sol_matter_full_ti=10} and Fig.~\ref{fig:sol_matter_full_ti=100}. The damping due to the expansion of the universe can be neglected at late times.

\begin{figure}[H]
\centering
\includegraphics[scale=1]{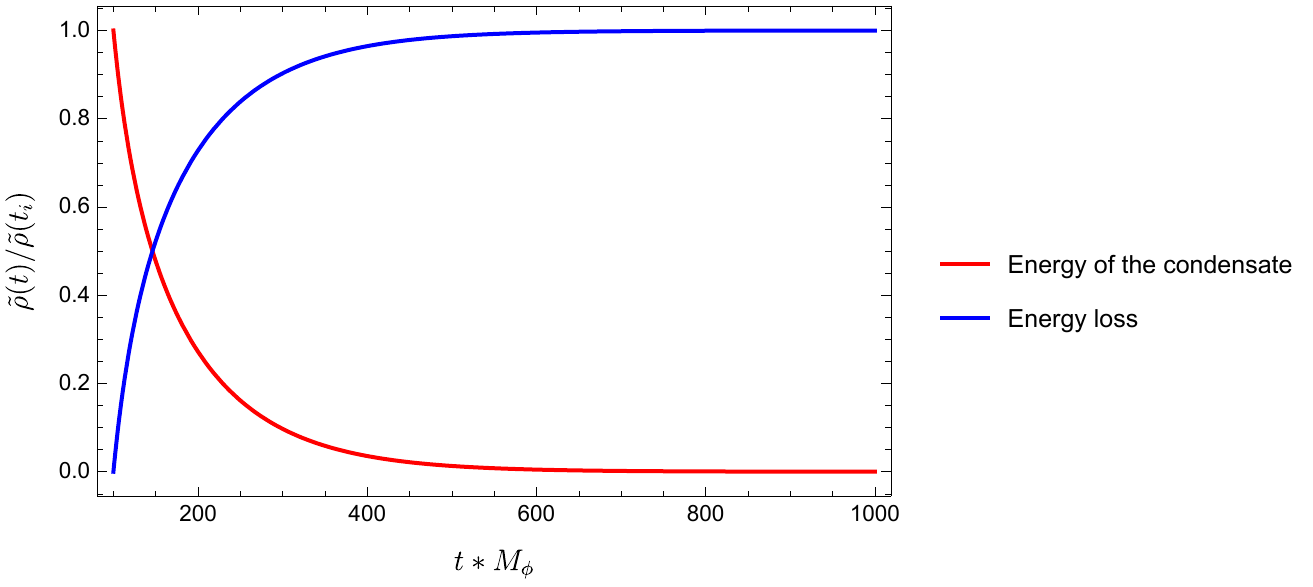}
\caption{Energy transfer from the condensate to the produced particles in a matter-dominated universe. The evolution of the condensate comoving energy density and of the energy loss from the condensate  are shown in the red and blue lines, respectively. The energy transfer from the condensate to the produced particles in a matter-dominated universe is efficient in presence of a non-vanishing self-energy correction.} 
\label{fig:energy_matter_full_ti=100}
\end{figure}

The energy density of the condensate in a matter-dominated universe can be approximated as
\begin{align}
    \rho_{\varphi}(t) \approx \frac{M_{\phi} ^2}{2\, t^2} \cdot \frac{\e ^{-2\gamma t}}{ c_0-2\, \sigma \,  \gamma \, \Gamma (-1,2 \gamma t)}\,,
\end{align}
which still satisfies Eq.~\eqref{eq:drho_dt_rad}, but with $A(t)$ given by Eq.~\eqref{eq:matter_full_A}. In the late-time limit $\gamma t \gg 1$, the comoving energy density of the condensate exponentially decreases with the decay rate $2 \gamma$. The energy transfer from the condensate to the produced particles in a matter-dominated universe is drawn in Fig.~\ref{fig:energy_matter_full_ti=100}. In presence of the $\gamma$ dissipation, the decay of the condensate comoving energy density is complete.

\end{appendix}

\bibliographystyle{utphys}
\bibliography{ref}{}

\end{document}